\documentclass[prx,showpacs,aps,twocolumn,amssymb,amsmath,notitlepage,superscriptaddress,longbibliography]{revtex4-1}

\usepackage{amsfonts}
\usepackage{graphicx,graphics,epsfig,times,bm,bbm,mathrsfs} 
\usepackage[normalem]{ulem}
\usepackage{subfigure}
\usepackage[pdfstartview=FitH]{hyperref}
\hypersetup{
    colorlinks=true,       
    linkcolor=cyan,          
    citecolor=magenta,        
    filecolor=magenta,      
    urlcolor=cyan,           
    runcolor=cyan
}
\usepackage{upgreek}

\newcommand{\beq}{\begin{equation}}
\newcommand{\eeq}{\end{equation}}
\newcommand{\beqnn}{\begin{equation*}}
\newcommand{\eeqnn}{\end{equation*}}
\newcommand{\beann}{\begin{eqnarray*}}
\newcommand{\eeann}{\end{eqnarray*}}

\newcommand{\bes} {\begin{subequations}}
\newcommand{\ees} {\end{subequations}}
\newcommand{\bea} {\begin{eqnarray}}
\newcommand{\eea} {\end{eqnarray}}

\newcommand{\ignore}[1]{}

\newcommand{\braket}[2]{\langle #1 | #2\rangle}
\newcommand{\ket}[1]{ | #1\rangle}
\newcommand{\bra}[1]{\langle #1 | }

\def\up{\uparrow}
\def\down{\downarrow}

\begin{document}

\title{Non-Stoquastic Interactions in Quantum Annealing via the Aharonov-Anandan Phase}

\author{Walter Vinci}
\affiliation{Department of Electrical Engineering, University of Southern California, Los Angeles, California 90089, USA}
\affiliation{Department of Physics and Astronomy, University of Southern California, Los Angeles, California 90089, USA}
\affiliation{Center for Quantum Information Science \& Technology, University of Southern California, Los Angeles, California 90089, USA}
\author{Daniel A. Lidar}
\affiliation{Department of Electrical Engineering, University of Southern California, Los Angeles, California 90089, USA}
\affiliation{Department of Physics and Astronomy, University of Southern California, Los Angeles, California 90089, USA}
\affiliation{Center for Quantum Information Science \& Technology, University of Southern California, Los Angeles, California 90089, USA}
\affiliation{Department of Chemistry, University of Southern California, Los Angeles, California 90089, USA}

\begin{abstract}

We argue that a complete description of quantum annealing (QA) implemented with continuous variables must take into account the non-adiabatic Aharonov-Anandan geometric phase that arises when the system Hamiltonian changes during the anneal. We show that this geometric effect leads to the appearance of non-stoquastic terms in the effective quantum Ising Hamiltonians that are typically used to describe QA with flux-qubits. We explicitly demonstrate the effect of these geometric interactions when QA is performed with a system of one and two coupled flux-qubits.
The realization of non-stoquastic Hamiltonians has important implications from a  computational complexity perspective, since it is believed that in many cases QA with stoquastic Hamiltonians can be efficiently simulated via classical algorithms such as Quantum Monte Carlo. It is well-known that the direct implementation of non-stoquastic interactions with flux-qubits is particularly challenging. Our results suggest an alternative path for the implementation of non-stoquastic interactions via geometric phases that can be exploited for computational purposes.

\end{abstract}
\maketitle


\textit{Introduction}.---%
It is well known that the solution of computational problems can be encoded into the ground state of a time-dependent quantum Hamiltonian. This approach is known as adiabatic quantum computation (AQC)~\cite{farhi2000quantum,farhi2001quantum,vanDam:01}, and is universal for quantum computing \cite{aharonov_adiabatic_2007} (for a review of AQC see Ref.~\cite{albash2016adiabatic}). 
Quantum annealing (QA) is a framework that incorporates algorithms~\cite{kadowaki_quantum_1998,Santoro,RevModPhys.80.1061} and hardware \cite{brooke2001tunable,Dwave,Johnson:2010ys,Harris:2010kx,McMahon:2016aa} designed to solve computational problems via quantum evolution towards the ground states of final Hamiltonians that encode classical optimization problems, without necessarily insisting on universality or adiabaticity.

QA thus inhabits a regime that is intermediate between the idealized assumptions of universal AQC and unavoidable experimental compromises. Perhaps the most significant of these compromises has been the design of \emph{stoquastic} quantum annealers. A Hamiltonian $H$ is stoquastic with respect to a given basis if $H$ has only real nonpositive off-diagonal matrix elements in that basis, which means that its ground state can be expressed as a classical probability distribution \cite{bravyi2006complexity,bravyi2009complexity}. Typically, one chooses the  computational basis, i.e., the basis in which the final Hamiltonian is diagonal. The computational power of stoquastic Hamiltonians has been carefully scrutinized, and is suspected to be limited in the ground-state AQC setting ~\cite{albash2016adiabatic}. E.g., it is unlikely that ground-state stoquastic AQC is universal~\cite{farhi2016quantum}. Moreover, under various assumptions ground-state stoquastic AQC can be efficiently simulated by classical algorithms such as quantum Monte Carlo~\cite{bravyi2006complexity,bravyi2009complexity,Heim:2014jf,Bravyi:2016aa}, though certain exceptions are known~\cite{Hastings:2013kk,jarret2016adiabatic}. 


One is thus naturally motivated to consider a departure from the stoquastic setting. 
Indeed, this is the setting of proofs of the universality of AQC and of various specific results that use non-stoquasticity to improve upon the performance of a stoquastic Hamiltonian \cite{Farhi:2011:QAA:2011395.2011396,Seoane:2012uq,crosson2014different,Seki:2015,Zeng:2016bs,Hormozi:2016aa}. For example, it is known that non-stoquastic interactions that are turned on temporarily during QA can modify the annealing path so that it encounters a polynomially small gap rather than an exponentially small one, by replacing a first order quantum phase transition by a second order one \cite{Nishimori:2016aa}. 

Introducing non-stoquastic interactions is especially important in the physical implementation of QA devices, in order to allow them to escape the trap of efficient classical simulability. As a case in point, and setting aside heavily debated concerns about whether such devices are sufficiently quantum \cite{SSSV,q-sig2,Albash:2014if,DWave-entanglement,Boixo:2014yu,Albash:2015pd}, despite intense efforts~\cite{q108,speedup,Hen:2015rt,Martin-Mayor:2015dq,King:2015zr,Boixo:2014yu,2016arXiv160401746M,Vinci:2016tg} there is currently no evidence of an example where stoquastic QA hardware such as the D-Wave devices~\cite{Dwave,Johnson:2010ys,Harris:2010kx} delivers a quantum speedup over all possible classical algorithms. This is true even though in this setting QA is not limited to ground state evolution. However, the implementation of non-stoquastic interactions is technologically challenging, at least with superconducting flux-qubits. The D-Wave devices, for example, have a scalable design that can only implement the Hamiltonian of the quantum transverse field Ising model, which is stoquastic. 
To remedy this would require additional couplings between the flux qubits, to realize, e.g., $\sigma^x\otimes\sigma^x$ interactions, in addition to the existing $\sigma^z\otimes\sigma^z$ interactions. This greatly complicates the design of the current generation of superconducting circuits.

Rather than attempt to introduce non-stoquasticity via new components, here we revisit the assumptions that lead to the derivation of the Hamiltonian generating the effective time evolution of a continuous-variable system, such as inductively coupled flux qubits. We show that non-stoquastic terms arise naturally as a non-adiabatic geometric phase, due to the Aharonov-Anandan effect~\cite{aharonov1987phase,ANANDAN1988171}. In other words, \emph{non-stoquastic terms are in fact present all along when QA is performed in systems of inductively coupled flux qubits}. We study these geometric terms in detail, and show that the geometric effect is amplified in proportion to the inverse gap of the flux Hamiltonian, appearing and then disappearing during the anneal. The geometric effect can thus be considered as a type of non-stoquastic catalyst~\cite{albash2016adiabatic}, with the potential to lead to quantum speedups \cite{Farhi:2011:QAA:2011395.2011396,Seoane:2012uq,crosson2014different,Seki:2015,Zeng:2016bs,Hormozi:2016aa,Nishimori:2016aa}. 
%
The presence of geometric terms also has clear experimental consequences. Geometric effects should be taken into account in the validation of current and future QA devices. Conversely, the same devices could be used to perform experimental measurements of non-trivial geometric phases. 

\textit{General formalism and the geometric term}.---%
For concreteness, we assume that QA is performed by implementing a time-dependent Hamiltonian that can be generically expressed as follows:
\beq
H(\bm{\phi},t) = -\frac12E_\text{C}\sum^n_{i=1}\partial^2_{\phi_i}  + P(\bm{\phi},t)\,.
\label{eq:H_ann}
\eeq
Such Hamiltonians can be realized with superconducting flux-qubits~\cite{RevModPhys.73.357}, in which case the continuous variables $\bm{\phi} = \{\phi_i\}_{i = 1}^n$ are the magnetic fluxes trapped by the $n$ flux-qubits, and $E_\text{C}$ represents a charging energy. The term $P(\bm{\phi},t)$ is a time-dependent potential that controls both the anneal and the interactions between qubits. We assume that the lowest $2^n$ energy levels of the Hamiltonian~\eqref{eq:H_ann} are separated from the rest of the spectrum by an energy gap.  At sufficiently low temperatures and for a slowly variable potential $P(\bm{\phi},t)$, the system of Eq.~\eqref{eq:H_ann} is then effectively confined to an $N=2^n$ dimensional Hilbert space $\mathcal H_N(t)$.

To proceed, we follow the approach introduced by Anandan in the discussion of non-adiabatic non-Abelian geometric phases~\cite{ANANDAN1988171}. We first consider the time-evolved $N$-dimensional subspace $\mathcal H_N(t)$, which is defined by the map 
\beq
U(t): \mathcal H_N(0) \mapsto \mathcal H_N(t)=U(t) \mathcal H_N(0),
\label{eq:H_evol}
\eeq
where $U(t) = \mathcal T \exp\left(-i \int_0^t H(t') dt' \right)$ is the unitary time-evolution operator ($\mathcal T$ denotes time ordering), and $H(t)$ is the Hamiltonian~\eqref{eq:H_ann}. 
Let $\{\ket{\psi_a(0)}\}_{a=1}^N$ denote an orthonormal basis for $\mathcal H_N(0)$. Thus, $\mathcal H_N(t)$ has a basis whose orthonormal elements 
obey the Schr{\"o}dinger equation:
\beq
 i \partial_t\ket{ \psi_a(t) } = H(t) \ket{ \psi_a(t) }.
 \label{eq:Schro}
 \eeq
We now define another (arbitrary) orthonormal basis $\ket{\psi'_a(t)}$ for $\mathcal H_N(t)$, such that $\ket{\psi_a(0)}= \ket{\psi'_a(0)}$. Then there exists an $N\times N$ unitary transformation $W(t)$ between the two bases such that $\ket{ \psi_b(t)} = \sum_{a=1}^N W_{ab}(t)  \ket{\psi'_a(t)}$.
An arbitrary state $\ket{\omega(0)} = \sum_a \omega_a(0)  \ket{\psi_a(0)}\in\mathcal H_N(0)$  thus evolves according to:
$\ket{\omega(t)} = \sum_a \omega_a(0)  \ket{\psi_a(t)} 
 = \sum_a \omega_a(t)  \ket{\psi'_a(t)}$, where 
 $\omega_a(t) \equiv  \sum_b W_{ab}(t) \omega_b(0)$.
The unitary  $W(t)$ can thus be considered as describing the effective evolution of the state $\ket{\omega(t) }$ inside the subspace $\mathcal H_N(t)$ in the frame rotating with the basis $\ket{\psi'_a(t)}$ (we derive the evolution equation of this basis in the SM, section~\ref{app:psi'-evolution}). 
%
By substituting 
the basis transformation into Eq.~\eqref{eq:Schro} one easily finds:
\beq
i \partial_t{W}(t) = {H}^{\text{eff}}(t){W}(t)\ , \quad {H}^{\text{eff}}(t) \equiv \tilde{H}(t)-G(t)\ ,
\eeq
where $\tilde{H}(t)$ and $G(t)$ are the $N\times N$ matrices defined by the following matrix elements, respectively:
\bes
\label{eq:7}
\begin{align}
\label{eq:7a}
\tilde H_{ab}(t)  &\equiv \bra{\psi'_a(t)}   H(t)  \ket{\psi'_b(t)}\,, \\
G_{ab}(t)  &\equiv  \bra{\psi'_a(t)}  i \partial_t \ket{\psi'_b(t)}\,. 
 \label{eq:E_and_G}
\end{align}
\ees
Let us now assume that $H(t)$ depends on $t$ only via the invertible and differentiable ``annealing schedule" $s\equiv \kappa^{-1}(\tau)$, where $\tau\equiv t/t_f$, and $t_f$ denotes the final time. Then it follows directly from Eq.~\eqref{eq:E_and_G} that $G(t)dt = G(s)ds$ (see the SM, section~\ref{app:A}), which  shows that $G(s)$ is a \emph{geometric} term, i.e., it depends only on the schedule $s$ (and not on its parametrization)~\cite{berry1984quantal,wilczek1984appearance}. Consequently, $W(t_f) = \mathcal T \exp\left[-i \int_0^1 H^{\text{eff}}(s)  ds \right]$ with the dimensionless effective Hamiltonian
\begin{align}
H^{\text{eff}}(s) = t_f \dot{\kappa}(s) \tilde{H}(s) - G(s) \ ,
\label{eq:H_eff}
\end{align}
where from hereon a dot denotes $d/ds$, and we set  $s\equiv\tau$ for simplicity. Note that the geometric term is negligible only in the adiabatic limit $t_f\rightarrow\infty$. As we shall see, is responsible for the appearance of non-stoquastic terms when $t_f$ is finite.


\begin{figure*}[t]
\subfigure[]{\includegraphics[width=1\columnwidth]{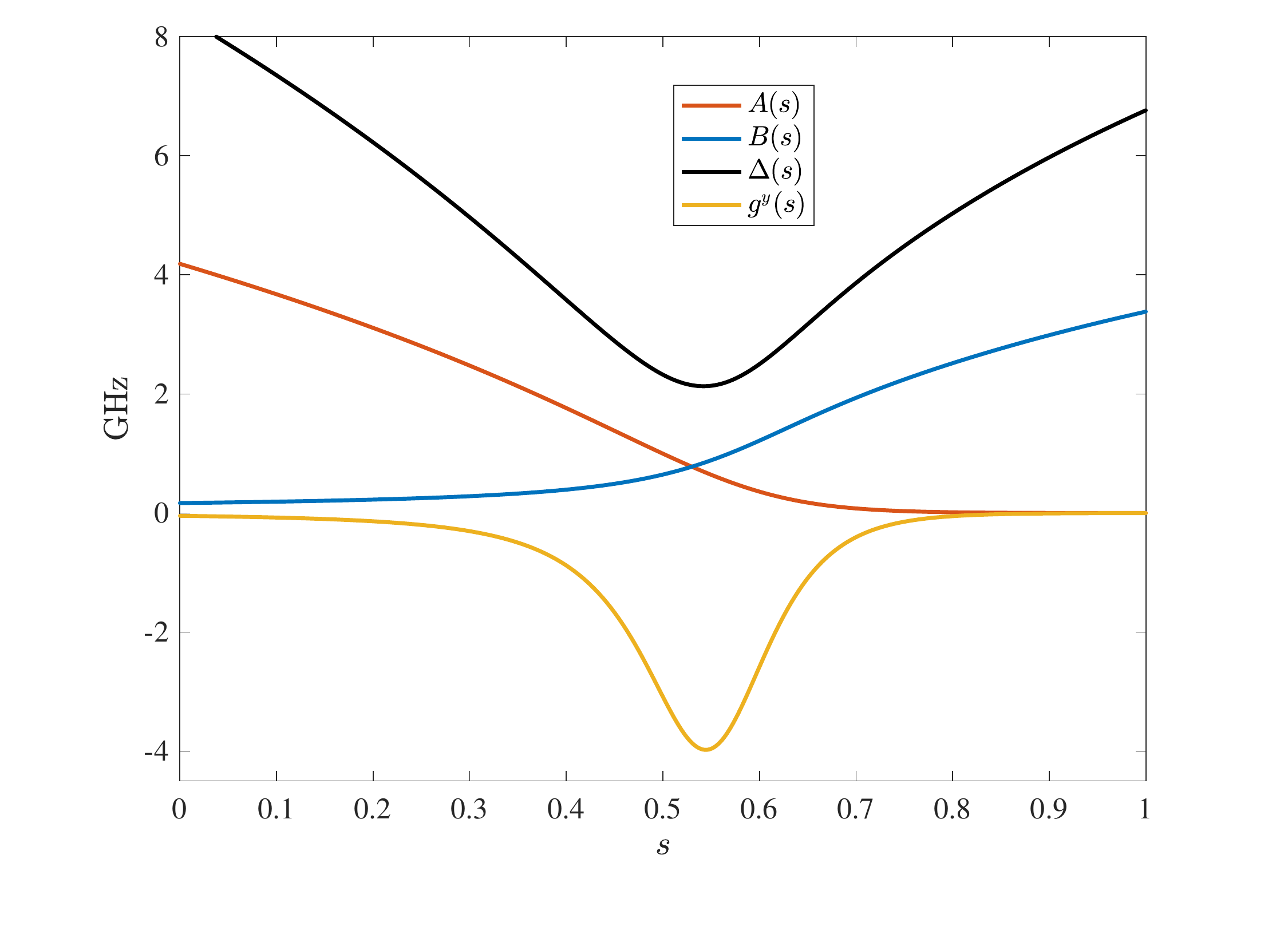} \label{fig:1a}}
\subfigure[]{\includegraphics[width=1\columnwidth]{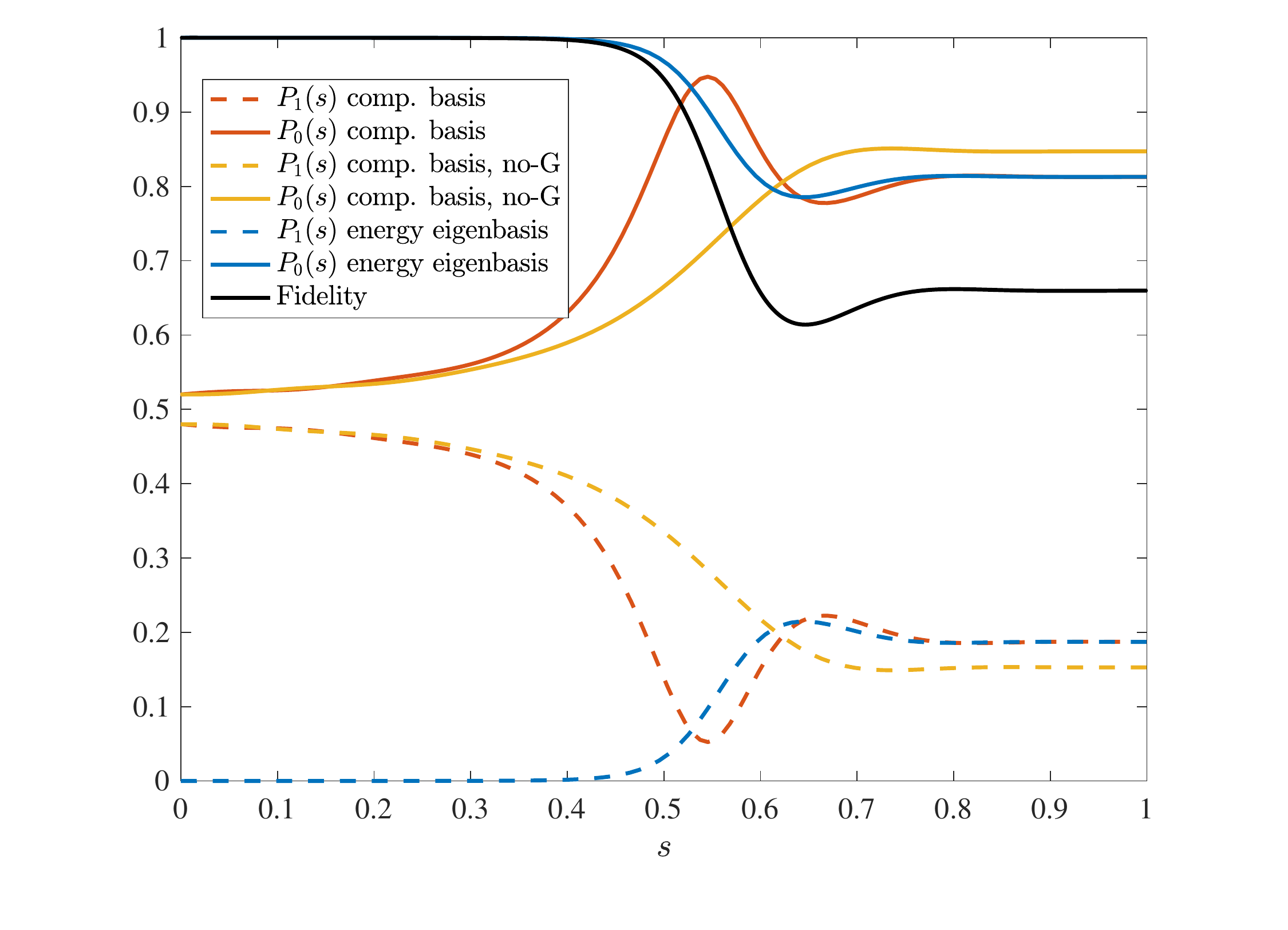} \label{fig:1b}}
\caption{One qubit. (a)  Annealing schedules $A(s)$ (red) and $B(s)$ (blue), gap $\Delta(s)$ (black) and the geometric term $g^y(s)$ (yellow) of  Eq.~\eqref{eq:H_1_rot}.  The functions $A(s)$ and $B(s)$ are obtained according to the procedure described in the SM, section~\ref{app:E}. (b) Populations of the states that are connected to the final ground state (solid) and final excited state (dashed), for $t_f=5$ (ns$/2\pi$). Blue and red lines represents the populations of the two states in the instantaneous energy eigenbasis [Eq.~\eqref{eq:H_1}] and the computational basis [Eq.~\eqref{eq:H_1_rot}], respectively. Because the two Hamiltonians are equivalent up to a time-dependent basis change, and the final basis is the computational basis in both cases, the populations at the end of the evolution ($s=1$) coincide. Yellow lines represent the populations during the anneal if the geometric term of Eq.~\eqref{eq:H_1_rot}  is dropped. The black line is the fidelity $|\braket{\psi^{\text{C}}_{1,\text{G}}(s)}{\psi^{\text{C}}_{1,\text{no-G}}(s)}|^2$ between the time-evolved states with (`G' subscript) and without (`no-G' subscript) geometric interactions. 
Results shown are for $\phi^x_{\text{CJJ}}(s) = 2.9(1-s)+2.2 s$ and parameter values for the flux-qubit Hamiltonian that match those of the highly coherent flux-qubits studied in Ref.~\cite{yan2016flux}:  $E_\text{S}/{2\pi \hbar } = 3.03$GHz, $E_\text{J}/{2\pi\hbar  } = 86.2$GHz [the control fluxes $\phi^x_{\text{CJJ}}(s)$ and $\phi^x(s)$ are shown in the SM, Fig.~\ref{fig:5a}]. 
}
\label{fig:1} 
\end{figure*}

\textit{Application to QA with superconducting flux qubits}.---%
Let $\{\ket{a(s)}\}$ denote the $N$ lowest energy instantaneous eigenstates of the original Hamiltonian~\eqref{eq:H_ann}, i.e., $H(s)\ket{a(s)} = E_a(s) \ket{a(s)}$. We identify the earlier $\{\ket{\psi'_a(s)}\}$ with these eigenstates, so that $W(s)$ describes the unitary evolution in the subspace rotating with the instantaneous eigenbasis of $H(s)$. In this basis, the QA process is described by the  dimensionless effective Hamiltonian~\eqref{eq:H_eff} with:
\begin{align}
\tilde{H}_{ab}(s) \equiv E_a(s) \delta_{ab}\,, \quad  G_{ab}(s) \equiv  \bra{a(s)}  i \partial_s \ket{b(s)}\,.
 \label{eq:H_eff_diagonal}
\end{align}
In QA applications, the Hamiltonian~\eqref{eq:H_ann} is designed such that its $N=2^n$ lowest energy levels can be put into $1$-to-$1$ correspondence with the energy levels of a transverse field Ising model with  $n$ qubits. Thus, there exists a unitary $V(s)$ such that, up to a term proportional to the identity matrix~\cite{Boixo:2014yu}:
\beq
V(s) \tilde{H}(s) V^\dagger(s) = A(s)\tilde H^X +  B(s) \tilde H^Z\,,
\label{eq:comp_basis}
\eeq
where the profile functions $A(s)$ and $B(s)$ are particular to the qubit Hamiltonian (see SM, section~\ref{app:E}), $\tilde H^X= -\sum_{i=1}^n \sigma_i^x$ is the usual transverse-field driver, and 
\beq
\tilde H^Z = \sum_{i=1}^n  h_i \sigma_i^z + \sum_{i>j}^n  J_{ij} \sigma_i^z\sigma_j^z
\label{eq:ising}
\eeq
is the problem Hamiltonian whose ground state encodes the answer to the optimization problem of interest. The unitary $V(s)$ is the transformation between the energy eigenbasis and the computational basis $\ket{\psi^{\text{C}}_a(s)}$. Note that the geometric term transforms as a geometric connection~\cite{ANANDAN1988171} (see SM, section~\ref{app:B}):
\beq
 G^{\text{C}}(s) = V(s) G(s) V^\dagger(s) + i V(s) \dot{V}^\dagger(s)\,.
 \label{eq:tildeG'}
\eeq
Quantum annealing of a system of $n$ flux-qubits controlled by the Hamiltonian~\eqref{eq:H_ann} is then described by the following effective Hamiltonian in the computational basis:
\beq
H^{\text{eff,C}}(s) = t_f\left( A(s)\tilde H^X +  B(s) \tilde H^Z\right) - G^\text{C}(s)\,.
\label{eq:eff_H}
\eeq
The first term, proportional to $t_f$, is the usual Hamiltonian discussed in literature in the context of QA. The second term has a geometric origin and is non-vanishing for finite annealing times $t_f$. The geometric term is non-stoquastic. To see this, note that the original Hamiltonian~\eqref{eq:H_ann} is real, and thus there is a basis choice in which the energy eigenbasis states $\ket{a(s)}$ have only real amplitudes. Consequently, the geometric term $G(s)$ in Eq.~\eqref{eq:H_eff_diagonal} is then a purely imaginary Hermitian matrix. The transformed geometric term $G^{\text{C}}(s)$ [Eq.~\eqref{eq:tildeG'}] is also purely imaginary in the computational basis, since the basis change of Eq.~\eqref{eq:comp_basis} can be performed with a real unitary (i.e., orthogonal) matrix $V(s)$. Therefore, including only interactions up to two-body terms, the geometric term can be written in the most general form as follows:
\beq
G^{\text{C}}(s) =  \sum_ig^{y}_i(s)   \sigma_i^y +  \sum_{i\neq  j} g^{xy}_{ij}(s)   \sigma_i^x \sigma_j^y + g^{zy}_{ij}(s) \sigma_i^z \sigma_j^y\,. 
\label{eq:G'-general-2q}
\eeq

\begin{figure*}[t]
\subfigure[]{\includegraphics[width=1\columnwidth]{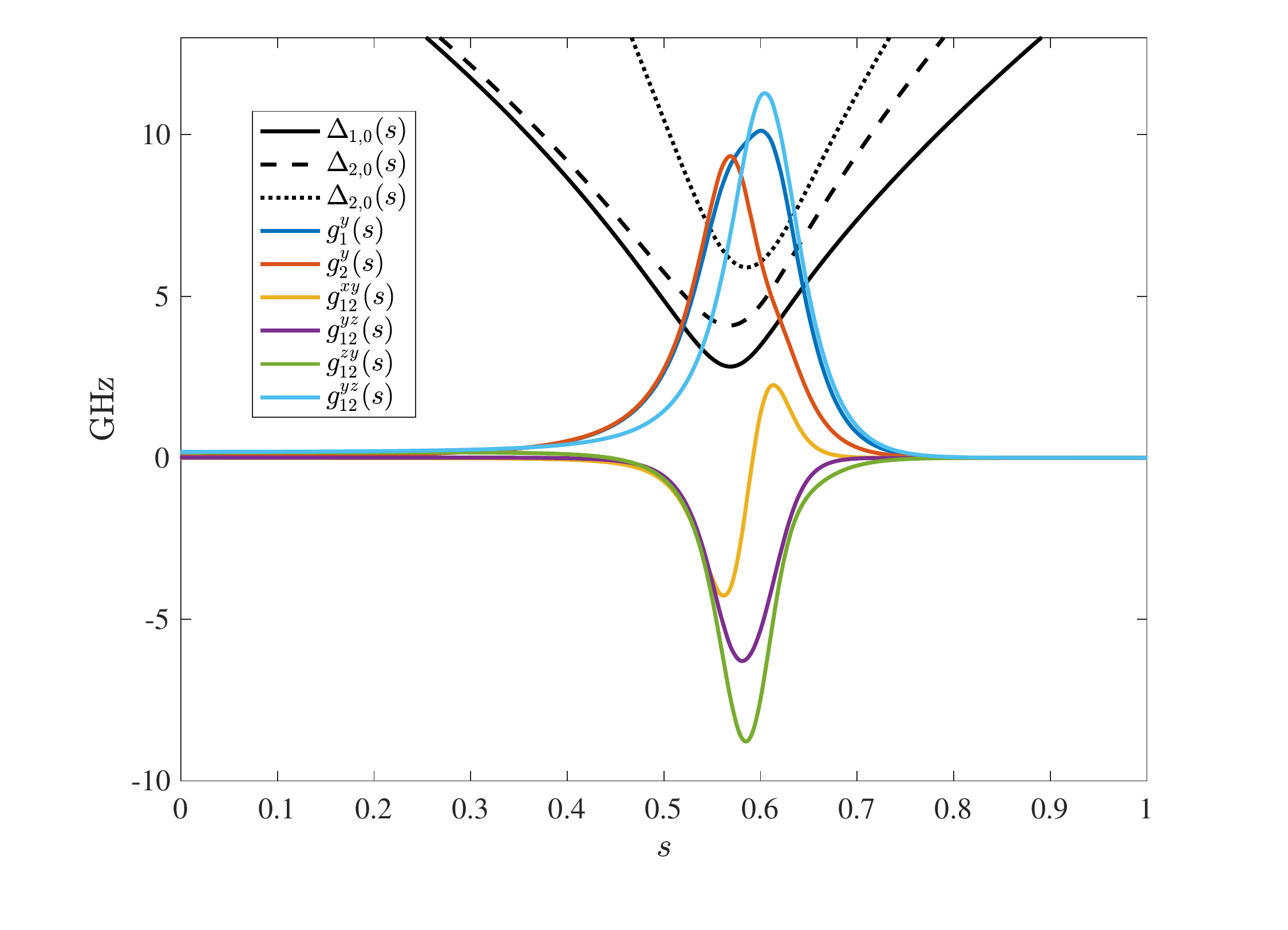} \label{fig:2a}}
\subfigure[]{\includegraphics[width=1\columnwidth]{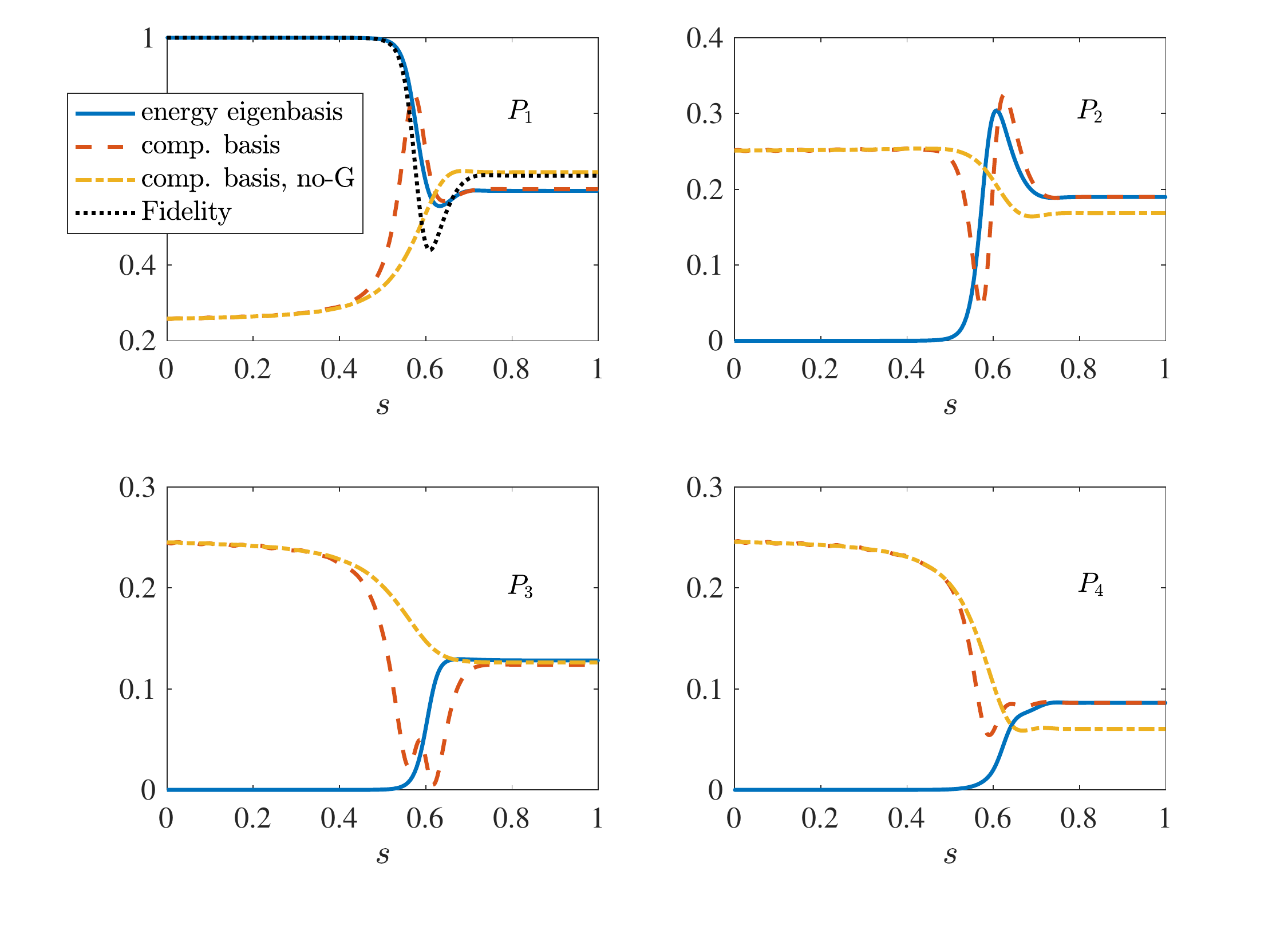} \label{fig:2b}}
\caption{Two interacting qubits. 
 (a) Gaps $\Delta_{k,0}(s)$ (black)  and the geometric terms $g^y_{i}(s), g^{xy}_{ij}(s), g^{zy}_{ij}(s)$ computed numerically using Eq.~\eqref{eq:H_two_qubit} for $h_1=1,h_2=0.4,J_{12}=-0.7$. (b) Populations of the states of the instantaneous basis that are connected to the final eigenstates, for the same value of the couplings as in (a) and $t_f=5$ (ns$/2\pi$). The black dotted line shown in the top left panel is the fidelity as explained in Fig.~\ref{fig:1}. 
Results shown are for $E_\text{C}/{2\pi \hbar } = 3.44$GHz,  $E_\text{J}/{2\pi\hbar  } = 684$GHz, $E_\text{L}/{2\pi  } = 570$GHz and $E_\text{M}/2\pi\hbar = 3.98$GHz, values that are typical for the D-Wave devices~\cite{Dwave,Johnson:2010ys,Harris:2010kx} [the control fluxes are shown in the SM, Fig.~\ref{fig:4a}]. For consistency with the one C-shunt flux qubit case we set $\phi^x_{\text{CJJ}}(s) = 2.6(1-s)+1.9 s$, while in the actual DW devices the schedule is chosen such that the field $\phi^x(s)$ is a linear function of $s$ [we have the same characteristic parametric relationship  $\phi^x(\phi^x_{\text{CJJ}})$].}
\label{fig:2} 
\end{figure*}

\textit{Quantum annealing with geometric terms: one qubit}.---%
We now apply the results obtained so far to QA with one qubit. For concreteness we focus on the C-shunt flux-qubit with three Josephson junctions. This qubit can be described by the following Hamiltonian~\cite{orlando1999superconducting,yan2016flux,2017arXiv170106544W} (see SM, section~\ref{app:D}): 
\bes
\label{eq:H_one_qubit}
\begin{align}
\label{eq:H_1_LL}
H_1(\phi,s) &= -\frac18 E_\text{S}\partial^2_\phi +  P(\phi,s)  \\
P(\phi,s) &= -2 E_\text{J}\left( \cos[\frac{1}{2}\phi^x_{\text{CJJ}}(s)]\cos[\phi^x(s) +2 \phi] +  \cos\phi  \right) \ ,
\label{eq:P}
\end{align}
\ees
where $\phi$ is the flux (in units of $\Phi_0/{2\pi}$) trapped in the superconducting ring,
$E_\text{S}$ is the charging energy of the shunting capacitor, and $\phi^x_{\text{CJJ}}(s)$ and $\phi^x(s)$ are external fluxes used to control the anneal.

We next construct the effective Hamiltonian of Eq.~\eqref{eq:eff_H}. We start by numerically computing the ground state $\ket{0(s)}$ and first excited state $\ket{1(s)}$ of the flux-qubit Hamiltonian Eq.~\eqref{eq:H_one_qubit} [examples are given in the SM, Figs.~\ref{fig:5b}-\ref{fig:5d}], from which we  numerically obtain the effective Hamiltonian [Eq.~\eqref{eq:H_eff}] in the instantaneous energy eigenbasis:
\beq
H_1^{\text{eff}}(s) =  t_f\frac{\Delta(s)}{2} \sigma^z -g(s)  \sigma^y\,,
\label{eq:H_1}
\eeq
(we have ignored a term proportional to the identity matrix) where $\Delta(s) \equiv E_1(s)-E_0(s)$ is the gap [plotted in Fig.~\ref{fig:1a}] and $g(s) \equiv  \bra{1(s)} \partial_s \ket{0(s)}$. We now transform to the computational basis using the unitary 
$V(s) = \exp\left[ \frac{i}{2} \arctan \left(\frac{A(s)}{B(s)}  \right) \sigma^y \right]$, after which
the Hamiltonian of Eq.~\eqref{eq:H_1} becomes
\beq
H_1^{\text{eff,C}}(s) =   t_f\left[A(s)\sigma^x  + B(s)  \sigma^z\right] -g^y(s)\sigma^y\, .
\label{eq:H_1_rot}
\eeq
This has the form of the most general single-qubit Hamiltonian: $H_1^{\text{eff,C}}(s) = \sum_{\alpha\in{x,y,z}} c_{\alpha}(s) \sigma^{\alpha}$.
It can be  checked easily that (see SM, section~\ref{app:C}):
\bes
\label{eq:18}
\begin{align}
\label{eq:Delta-A-B}
\Delta(s) &= 2\sqrt{A^2(s)+ B^2(s)}\,,  \\
g^y(s) &=  g(s) + \frac{2}{\Delta^2(s)}\left[\dot{A}(s)B(s) -  A(s) \dot{B}(s) \right]\ .
\label{eq:g^y}
\end{align}
\ees
By defining the computational basis as the basis of states with well-defined persistent current, the annealing schedule functions $A(s)$ and $B(s)$ [shown in Fig.~\ref{fig:1a}] can be determined in terms of the external fluxes $\phi^x_{\text{CJJ}}$ and $\phi^x$ of Eq.~\eqref{eq:P} (see the SM, section~\ref{app:E}). Thus, Eq.~\eqref{eq:g^y} shows that these flux parameters in turn control the properties of the geometric term $g^y$. The profile function $g^y(s)$ for the geometric term, also shown in Fig.~\ref{fig:1a}, is non-vanishing towards the middle of the adiabatic evolution, when the gap closes.  
The effects of the geometric term are shown in Fig.~\ref{fig:1b}, obtained by numerically solving for the corresponding unitary evolution. The effects become significant when the gap is small. There is also a significant effect on the final ground state population.

\textit{Quantum annealing with geometric terms: two qubits}.---%
To study the interacting case, we consider two inductively-coupled compound Josephson junction flux qubits~\cite{yan2016flux} (see SM, section~\ref{app:D}):
\bea
&  H_2(\phi_1,\phi_2,s,h_1,h_2,J_{12})  = H_1(\phi_1,s,h_1)+& \nonumber \\
& +   H_1(\phi_2,s,h_2) + P_{\text{int}}(\phi_1,\phi_2,s,J_{12})\,, &
\label{eq:H_two_qubit}
\eea
where the flux-qubit Hamiltonian is given by 
\bes
\label{eq:H_one_qubit_DW}
\begin{align}
\label{eq:H_1_DW}
&H_1(\phi,s,h_i) = - \frac14 E_\text{C}\partial^2_\phi +  P(\phi,s,h_i)  \\
&P(\phi,s,h_i) = 2E_\text{J}\cos(\phi)\cos\left[\frac{\phi^x_{\text{CJJ}}(s)}{2}\right]+E_\text{L}\frac{\left[\phi-h_i\phi^x(s)\right]^2}{2}
\end{align}
\ees
and the interaction potential is explicitly written as:
\beq
  P_{\text{int}}(\phi_1,\phi_2,s,J_{12})  = -J_{12} E_\text{M} [\phi_1-\phi^x(s)][\phi_2-\phi^x(s)]\ .\nonumber
\label{eq:H_int}
\eeq

Proceeding as in the single qubit case, we start from the Hamiltonian~\eqref{eq:H_two_qubit} to numerically compute $\tilde{H}$ and $G$ appearing in Eq.~\eqref{eq:H_eff_diagonal}, and construct the effective Hamiltonian in the instantaneous energy eigenbasis [Eq.~\eqref{eq:H_eff}].
Once again, the effective Hamiltonian can be expressed in the computational basis by (numerically) finding a unitary $V$ such that the final Hamiltonian reads:
\bea
\label{eq:H_2_rot}
&H_2^{\text{eff,C}}(s,h_1,h_2,J_{12}) =   t_fA(s) \left(\sigma_1^x + \sigma_2^x\right) +  &  \\
& +  t_fB(s) \left(h_1 \sigma_1^z+ h_2  \sigma_2^z+J_{12}\sigma_1^z\sigma_2^z\right) -G^{\text{C}}(s,h_1,h_2,J_{12})\, .& \nonumber 
\eea
This is the usual transverse field Ising model, plus an additional geometric term, whose general form is given in Eq.~\eqref{eq:G'-general-2q}. Figure~\ref{fig:2a} shows the gaps $\Delta_{k,0} = E_k-E_0$, where $\{E_k\}_{k=0}^3$ are the ordered eigenvalues of $H_2^{\text{eff,C}}$, and the components of the geometric term  $G^{\text{C}}(s,h_1,h_2,J_{12})$ in the computational basis as defined in Eq.~\eqref{eq:G'-general-2q}. Fig.~\ref{fig:2a} shows that, in general, the geometric functions $g^{y}_i(s), g^{xy}_{ij}(s),g^{zy}_{ij}(s)$ are all non-vanishing, and their magnitude grows as the ground state gap shrinks. In particular, as in the single-qubit case, geometric effects introduce a non-stoquastic contribution to the driver term which is non-vanishing towards the middle of the anneal. Fig.~\ref{fig:2b} shows the effect of the geometric interactions in the annealing of the system of two coupled qubits under consideration. As in the single qubit case, we see that ignoring the geometric terms results in consistently different final populations in the computational basis.

\begin{figure}[t]
\includegraphics[width=1\columnwidth]{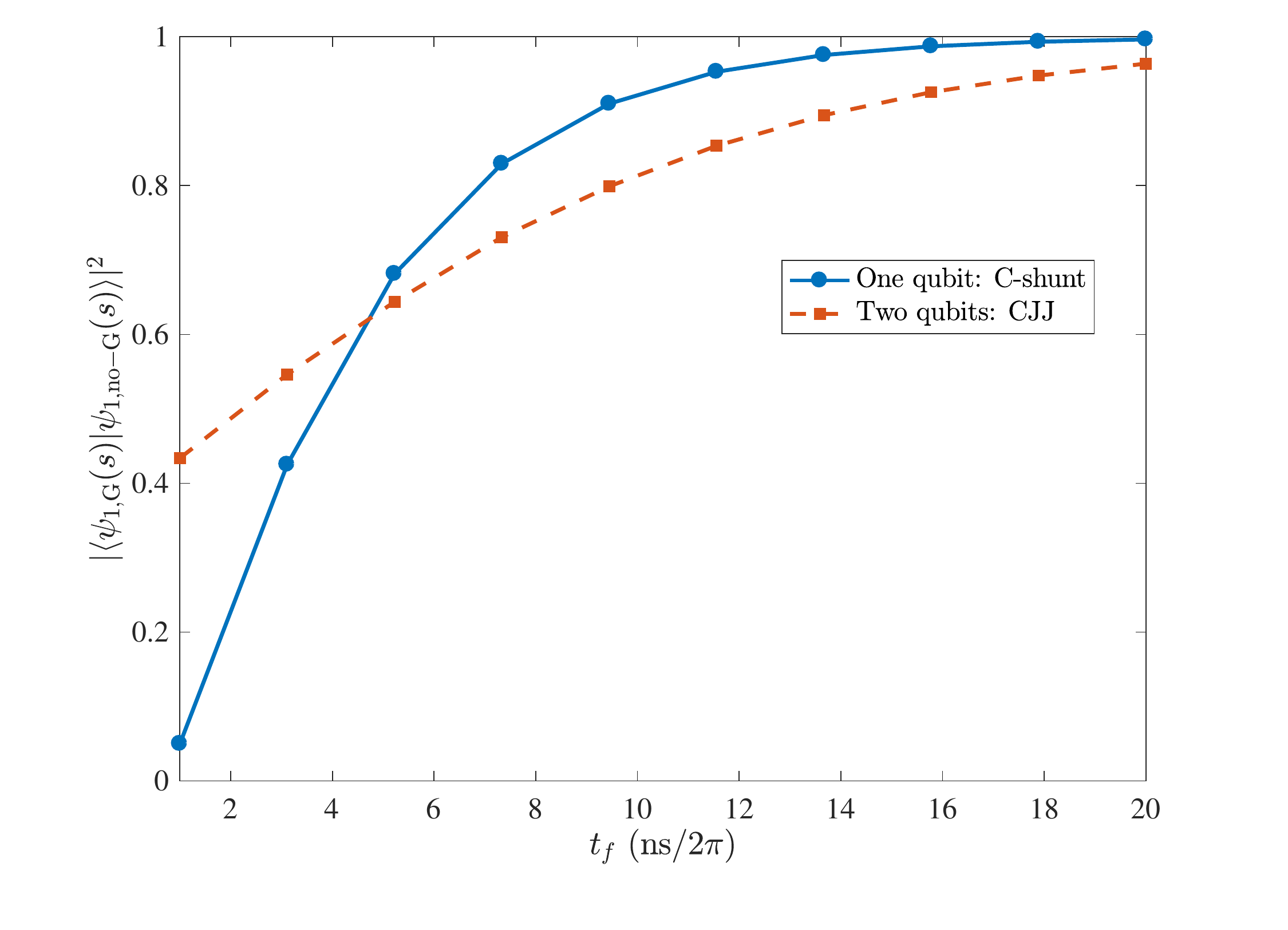}
\caption{Fidelity at the end of the anneal as a function of the annealing time for single qubit (blue) and two qubits (red). }
\label{fig:3} 
\end{figure}

\textit{Dependence on the annealing time}.---%
The contribution of the geometric terms is inherently a non-adiabatic effect, arising when diabatic transitions populate the excited states. In Fig.~\ref{fig:3} we show the fidelity $|\braket{\psi^{\text{C}}_{1,\text{G}}(s)}{\psi^{\text{C}}_{1,\text{no-G}}(s)}|^2$ between the time-evolved states with and without geometric terms, as a function of the total annealing time. As expected, the effect of the geometric terms increases with decreasing annealing time and it is reflected in the decreasing fidelity. Figure~\ref{fig:3} shows that the total annealing time over which this contribution is significant is on the order of a few nanoseconds, for parameters relevant for current QA devices. While this is significantly shorter than the typical microsecond timescale of current QA experiments using the D-Wave devices, this is the case for the one and two-qubit cases which we have analyzed here. Since, as is clear from Eq.~\eqref{eq:g^y} and from Fig.~\ref{fig:2}, the magnitude of the geometric term grows in inverse proportion to the gap, we expect it to become more significant for multi-qubit problems whose gap dependence can be inverse polynomial or even exponential. This effect is already visible in Fig.~\ref{fig:3}, which shows that the fidelity for the two-qubit system tends to be smaller than the one-qubit system for larger annealing times.

\textit{Conclusions}.---%
We have shown that even the simplest implementation of QA with flux-qubits induces effective Hamiltonians with non-stoquastic interactions arising from a geometric phase. The appearance of such interactions is ubiquitous when the Hamiltonian of a continuous-variable system is changed over time, and the evolution is non-adiabatic, due to the appearance of the Aharanov-Anandan effect.
Since arbitrarily small gaps are inevitable in QA for hard optimization problems, non-adiabatic evolutions are ultimately inescapable. We thus argue that, similarly, the geometric effects studied here are unavoidable and relevant in practical applications of QA. Moreover, since these geometric effects give rise to non-stoquastic terms in the effective Hamiltonian, they provide a natural and desirable mechanism for avoiding classically efficient simulation of QA. This may point to the possibility of ``quantum supremacy" experiments with QA devices featuring fewer than $100$ physical qubits ~\cite{preskill2012theory,boixo2016characterizing,Aaronson:2016aa,fefferman2017exact}.

\acknowledgements
We are grateful to Dr. Andrew Kerman for useful discussions. This work was supported under ARO grant number W911NF-12-1-0523, ARO MURI Grant Nos. W911NF-11-1-0268 and W911NF-15-1-0582, and NSF grant number INSPIRE-1551064.

\appendix

\section{Evolution equation of the $\{\ket{\psi'_a(t)}\}$ basis}
\label{app:psi'-evolution}

Recall that the $\{\ket{\psi_a(t)}\}$ basis was chosen to satisfy the Schr\"{o}dinger equation with the given Hamiltonian $H(t)$ [Eq.~\eqref{eq:Schro}]: $ \ket{ \dot{\psi}_a(t) } = -i H(t) \ket{ \psi_a(t)}$, where in this section dot denotes $\partial_t$. Here we derive the evolution equation satisfied by the $\{\ket{\psi'_a(t)}\}$ basis, which is related to the $\{\ket{\psi_a(t)}\}$ basis via the unitary $W(t)$:
\beq
\ket{\psi'_b(t)} = \sum_{a} [W^\dagger(t)]_{ab}  \ket{\psi_a(t)}\, .
\label{eq:U_base}
\eeq
Differentiation yields:
\begin{align}
\ket{\dot{\psi}'_b(t)} &= \sum_{a} [\dot{W}^\dagger]_{ab}  \ket{\psi_a(t)} + [W^\dagger(t)]_{ab}  \ket{\dot{\psi}_a} \notag \\
& = \sum_{a} \left([\dot{W}^\dagger]_{ab} -i [W^\dagger(t)]_{ab}  H(t) \right)\ket{ \psi_a(t)} \notag \\
& = \sum_{ac} \left([\dot{W}^\dagger]_{ab} -i [W^\dagger(t)]_{ab}  H(t) \right)W_{ca}(t)\ket{ \psi'_c(t)} \notag \\
& = \sum_{c} [W\dot{W}^\dagger]_{cb}\ket{\psi'_c(t)} -i H(t) \ket{\psi'_b(t)} \ .
\end{align}
Thus, the evolution equation satisfied by the basis element $\ket{\psi'_a(t)}$ is
\beq
\ket{\dot{\psi}'_a(t)}  = \sum_{b} i [\tilde{H}^{\text{eff}}(t)]_{ba}\ket{\psi'_b(t)} - i H(t) \ket{\psi'_a(t)} \ ,
\eeq
where we used $\dot{W} = -i {H}^{\text{eff}}(t){W}(t)$ and unitarity, and defined $\tilde{H}^{\text{eff}}(t) \equiv W^\dagger(t) {H}^{\text{eff}}(t) W(t)$.

\section{Proof that $G$ is a geometric term}
\label{app:A}

Let us show explicitly that $G$ is a geometric term, i.e., that $G(t)dt = G(s)ds$ where $s=\kappa^{-1}(t/t_f)$. Note that since we assumed that $\kappa$ is invertible and differentiable, we can write $t = t_f \kappa(s)$, so that $dt = t_f \frac{d\kappa(s)}{ds}ds$ and $\partial_t = \frac{ds}{dt}\partial_s = \frac{1}{t_f} \left[\frac{d\kappa(s)}{ds}\right]^{-1}\partial_s$. Thus 
\bes
\begin{align}
&G_{ab}(t)dt = \bra{\psi'_a(t)}  i \partial_t \ket{\psi'_b(t)} dt  \\
&\quad = \bra{\psi'_a(s)}  i \left[\frac{1}{t_f} \left(\frac{d\kappa(s)}{ds}\right)^{-1}\partial_s\right] \ket{\psi'_b(s)} t_f \frac{d\kappa(s)}{ds}ds \notag \\
&\quad = G(s) ds \ ,
\end{align}
\ees
where in the second line we used the assumption that $H(t)$ depends on $t$ only via $s$, which means, by Eq.~\eqref{eq:7a}, that the same must be true of $\psi'_a(t)$.

\section{Perturbative Derivation of Profile Functions $A(s)$ and $B(s)$}
\label{app:E}

\subsubsection{CJJ Flux-Qubit}

\begin{figure*}[t]
\subfigure[]{\includegraphics[width=1\columnwidth]{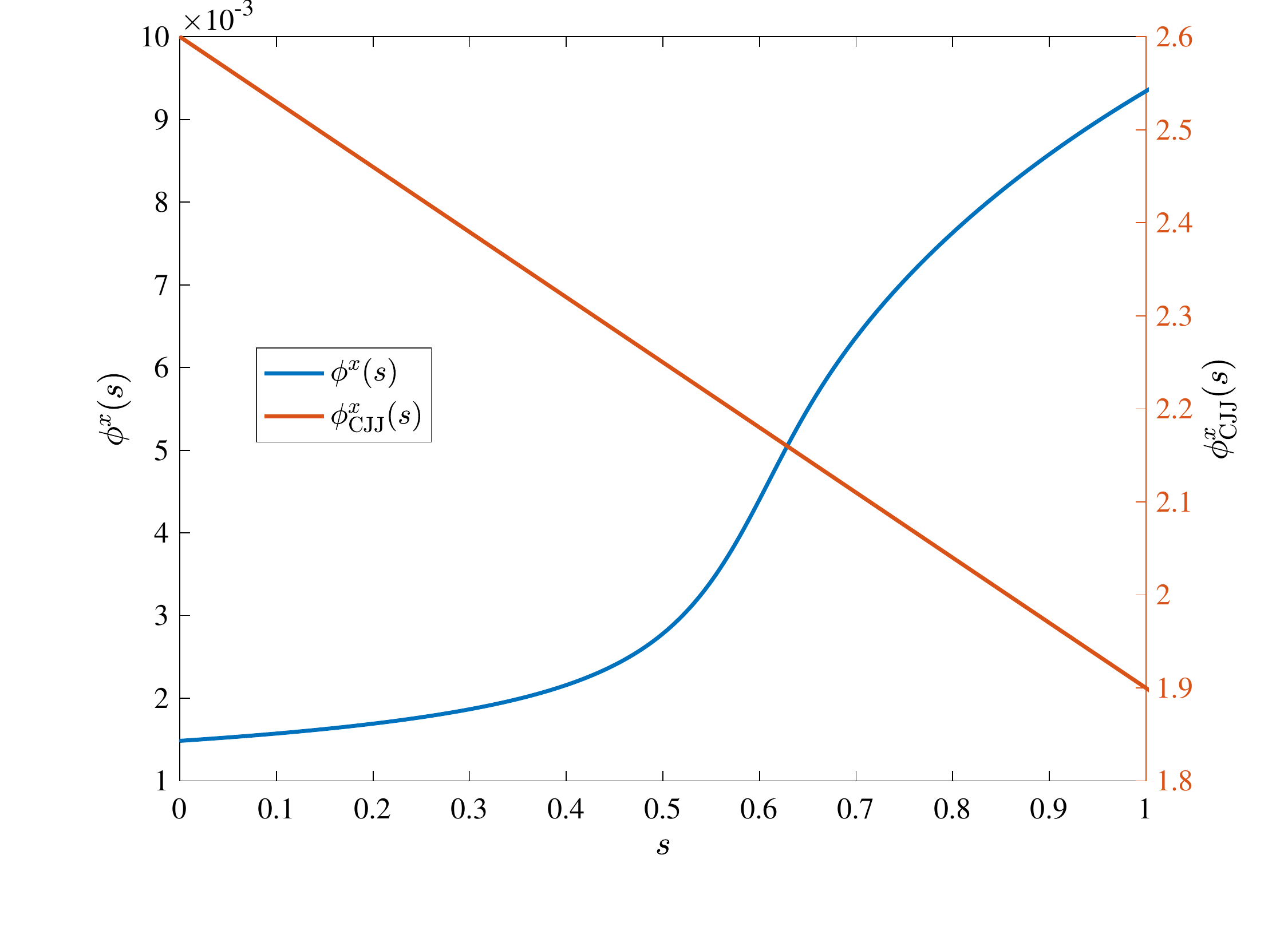} \label{fig:4a}}
\subfigure[] {\includegraphics[width=1\columnwidth]{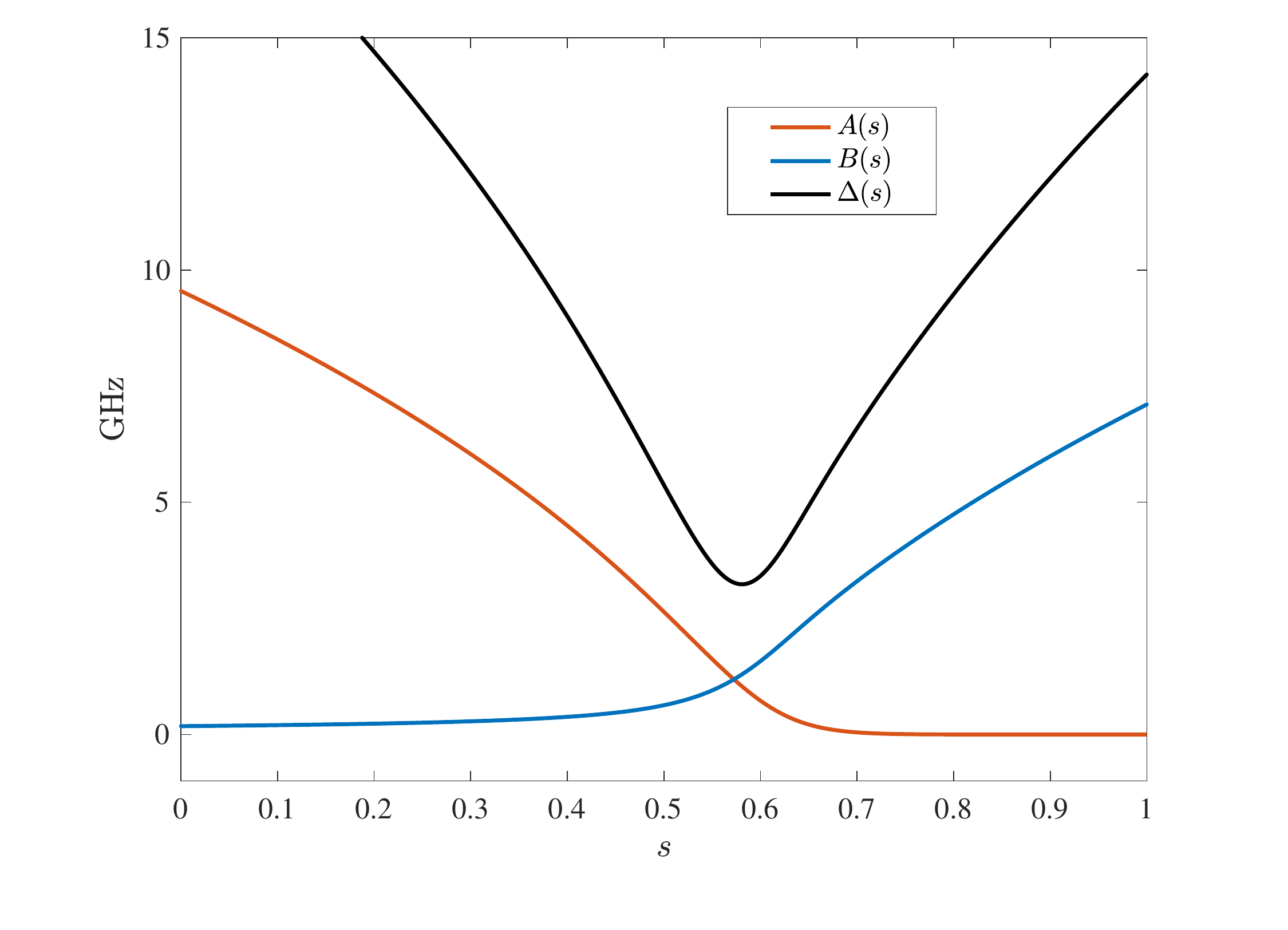} \label{fig:4b}}
\subfigure[] {\includegraphics[width=1\columnwidth]{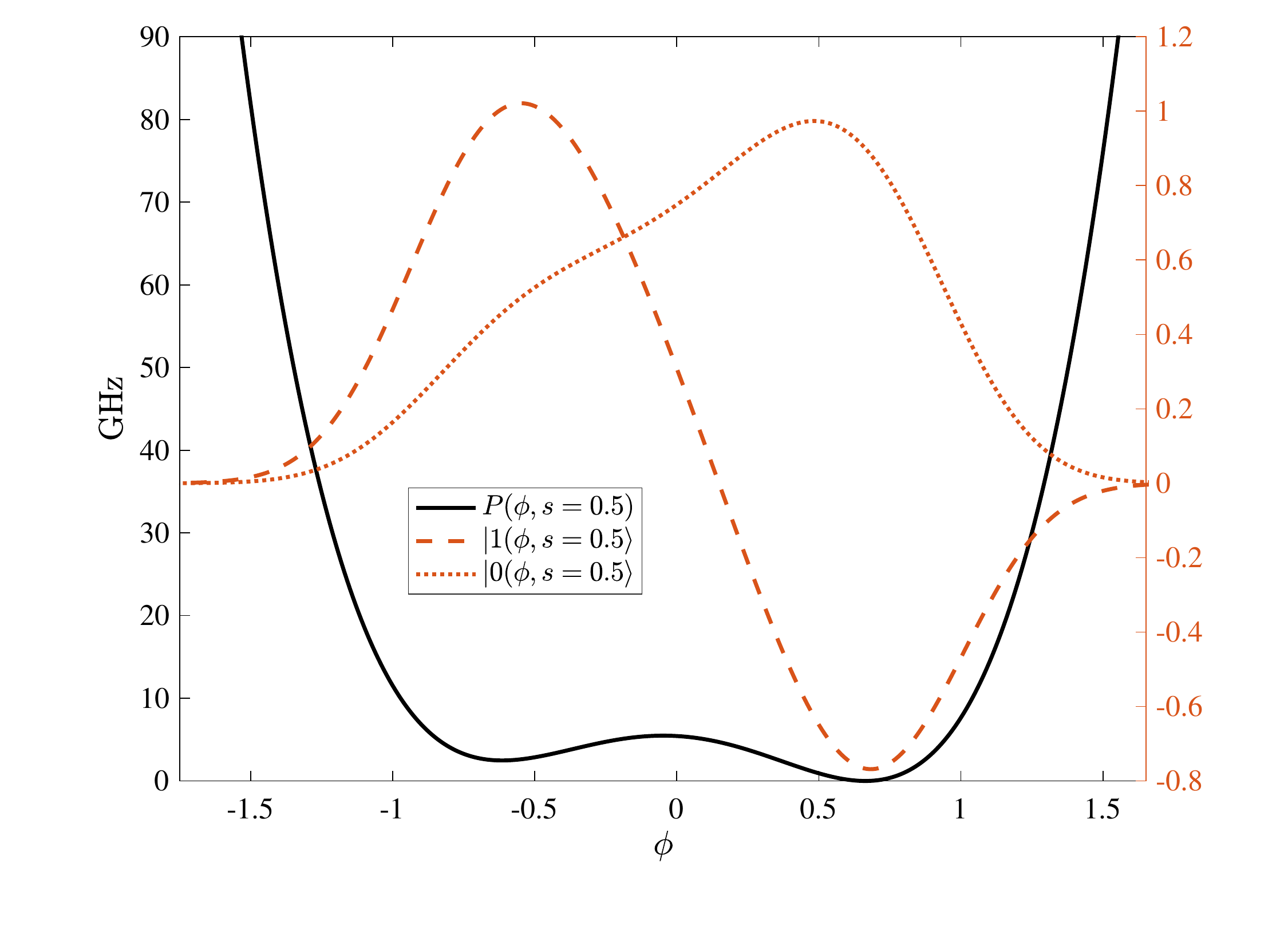} \label{fig:4c}}
\subfigure[] {\includegraphics[width=1\columnwidth]{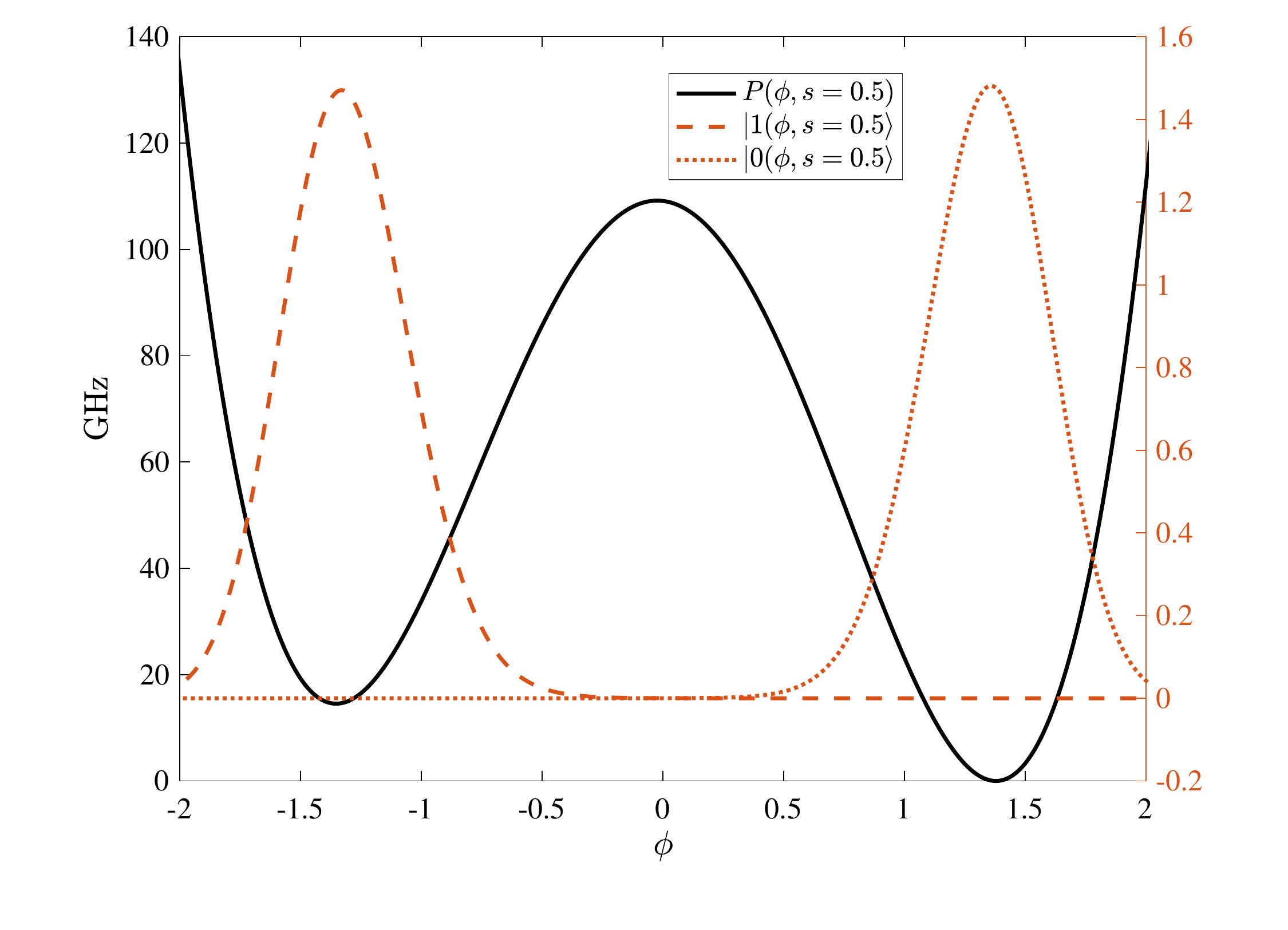} \label{fig:4d}}
\caption{CJJ flux-qubit Eq.~\ref{eq:H_one_qubit_DW}. (a)   The $s$ dependence of the flux $\phi_{\text{CJJ}}^x(s)$ (right axis) controls the annealing schedule. The flux $\phi^x(s)$ is computed accordingly. (b)   Gap and profile functions  $A(s)$ and $B(s)$. (d)  The potential $P(\phi,s)$ (black solid lines), and the corresponding wavefunctions of $\ket{0(s)}$ (dotted lines) and $\ket{1(s)}$ (dashed lines) for different values of the annealing parameter $s$: (c) $s=0.5$, and (d) $s=1$.}
\label{fig:4} 
\end{figure*}

\begin{figure*}[t]
\subfigure[]{\includegraphics[width=1\columnwidth]{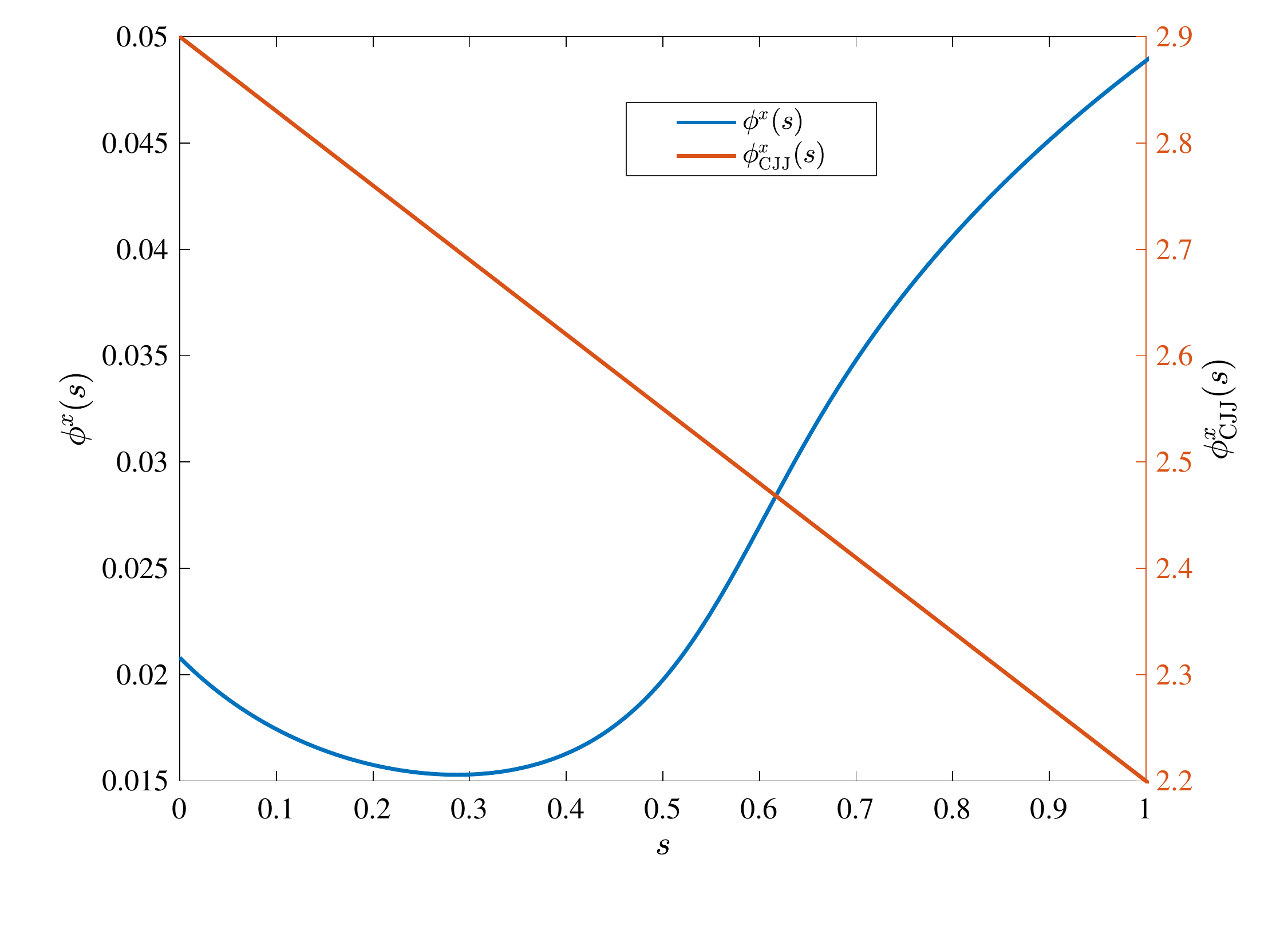} \label{fig:5a}}
\subfigure[] {\includegraphics[width=1\columnwidth]{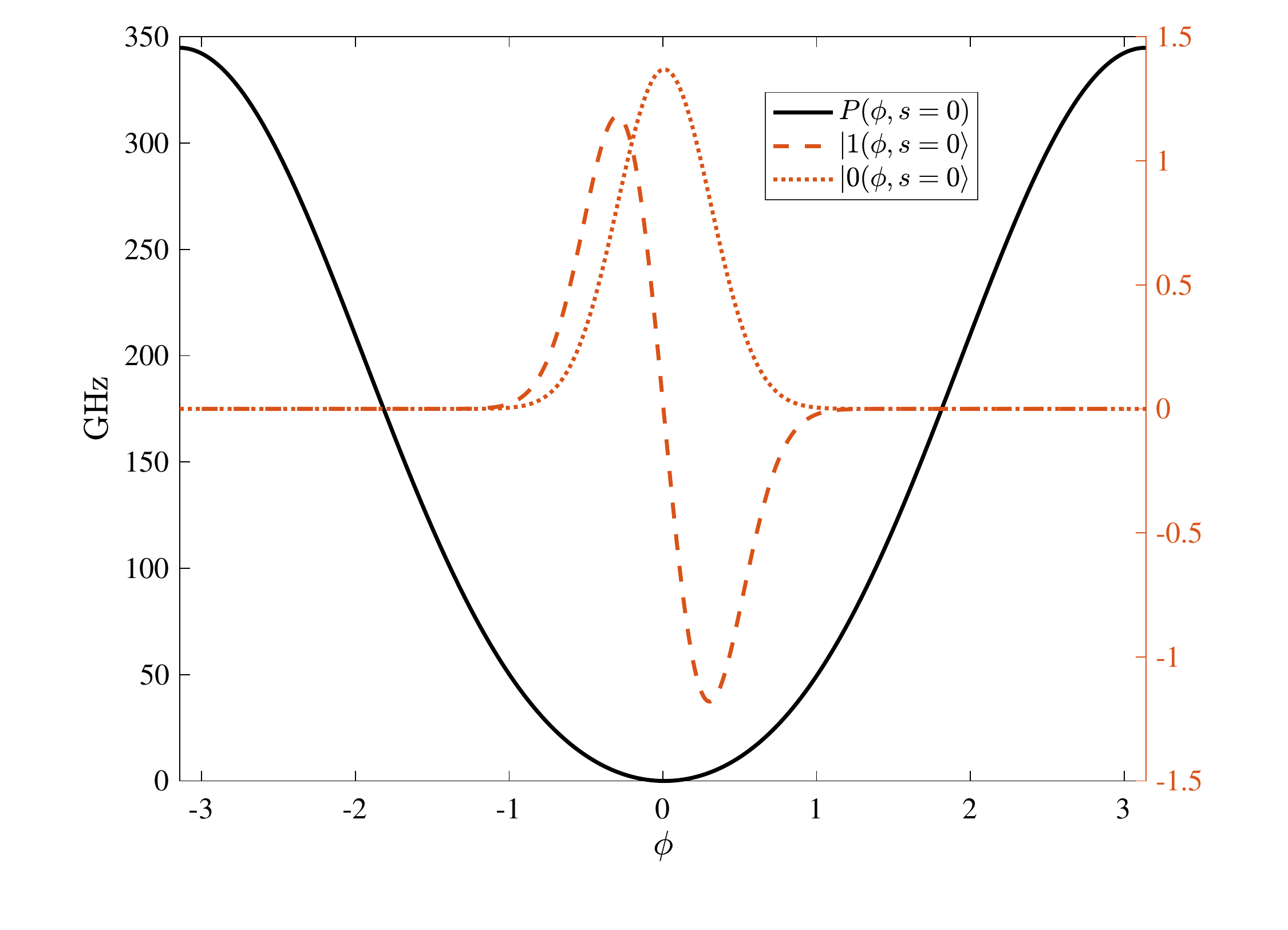} \label{fig:5b}}
\subfigure[] {\includegraphics[width=1\columnwidth]{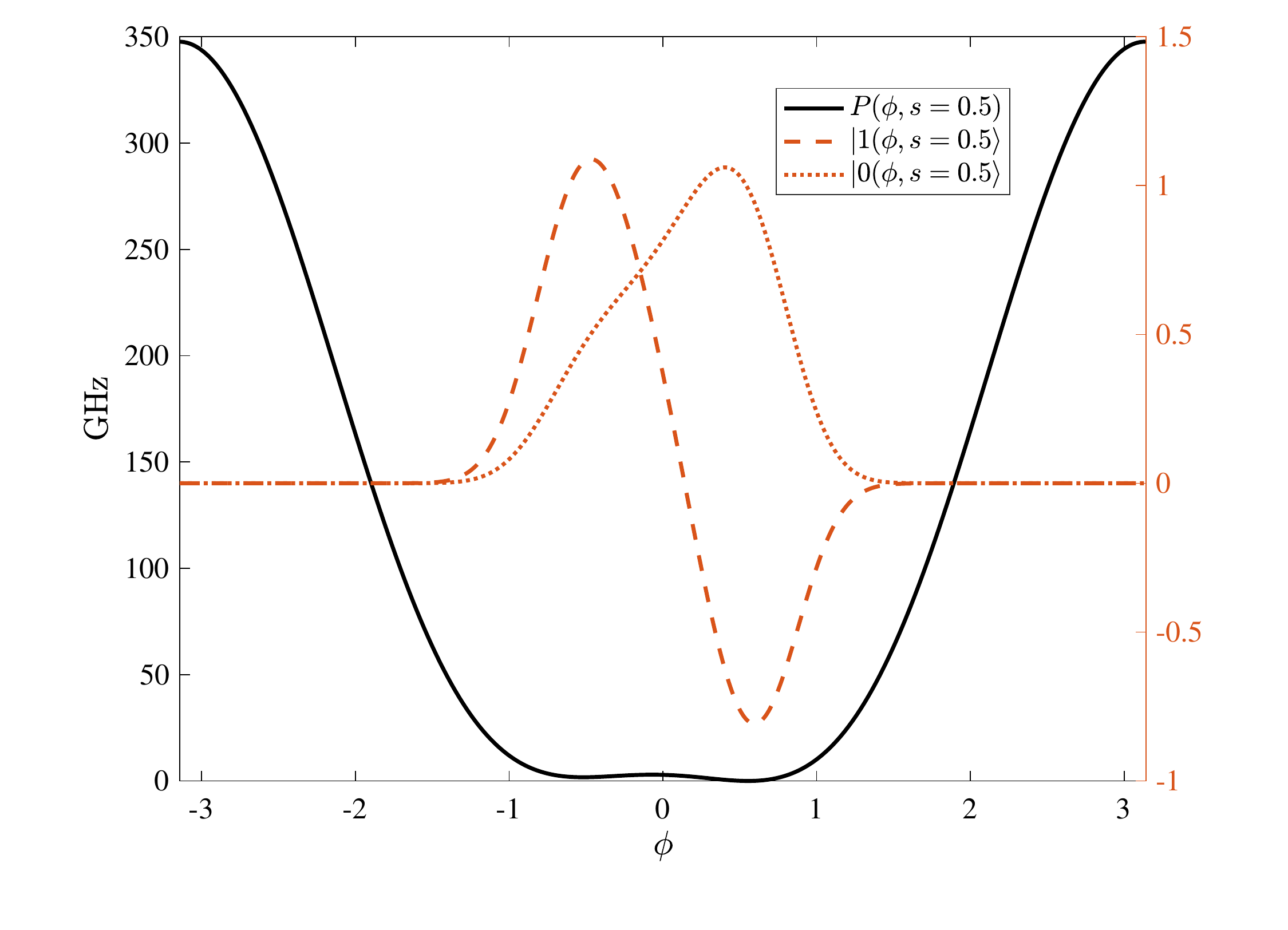} \label{fig:5c}}
\subfigure[] {\includegraphics[width=1\columnwidth]{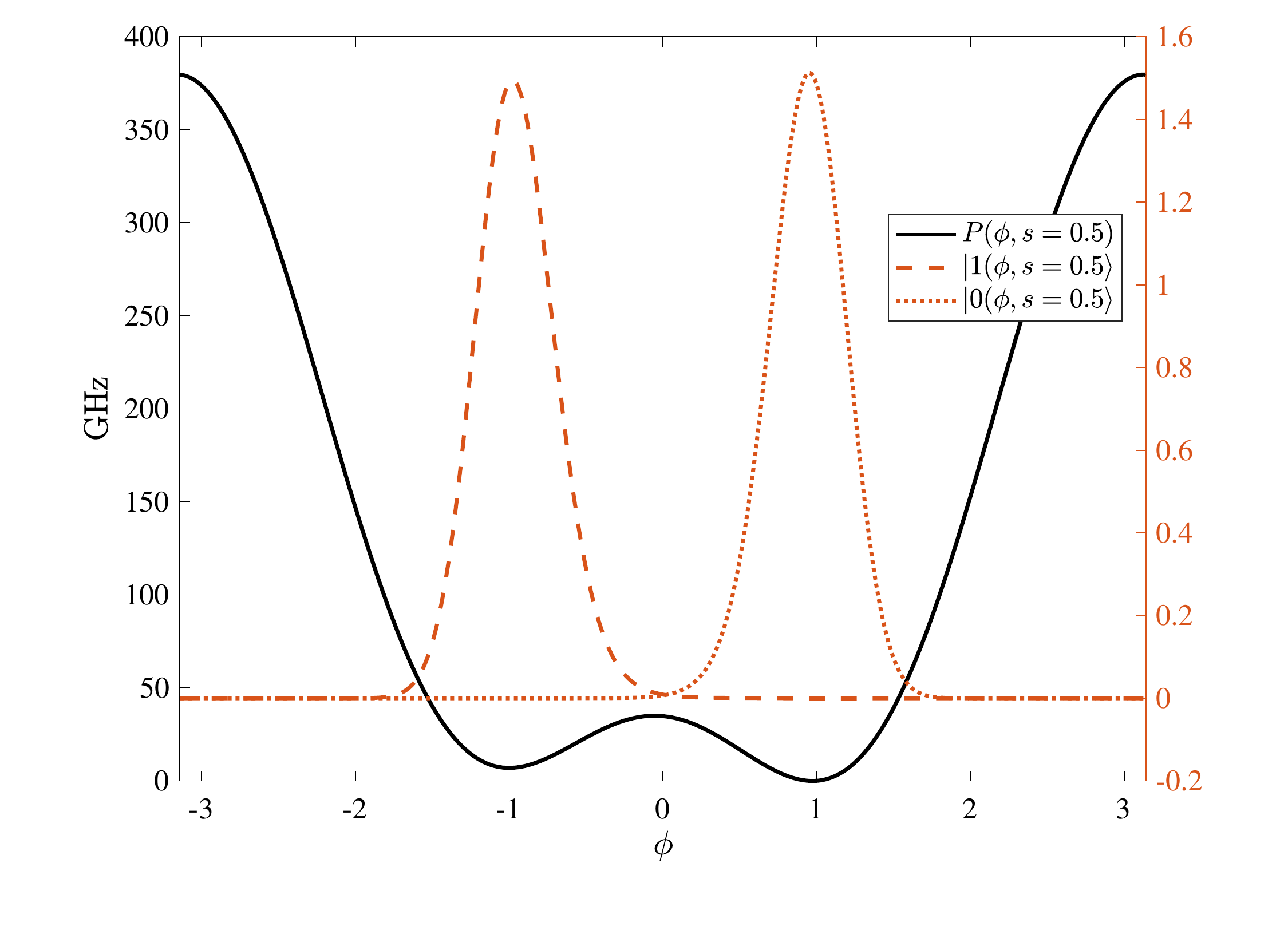} \label{fig:5d}}
\caption{ C-shunt flux-qubit Eq.~\ref{eq:H_one_qubit}. (a) The $s$ dependence of the flux $\phi_{\text{CJJ}}^x(s)$ (right axis) controls the annealing schedule. The flux $\phi^x(s)$ is computed accordingly. (b)-(d)  The potential $P(\phi,s)$ (black solid lines), and the corresponding wave functions of $\ket{0(s)}$ (dotted lines) and $\ket{1(s)}$ (dashed lines) for different values of the annealing parameter $s$: (b) $s=0$, (c) $s=0.5$, and (d) $s=1$.}
\label{fig:5} 
\end{figure*}

In this section we closely follow Ref.~\cite{Boixo:2014yu} to briefly describe how the annealing profile functions $A(s)$ and $B(s)$ can be calculated. The main observation is that the control flux  $\phi^x$, i.e., the bias between the two potential wells, is a small perturbation for the potential of $H_1$ in Eq.~\eqref{eq:H_one_qubit_DW}. The eigenstates $\ket{0(s)}$ and $\ket{1(s)}$ of the unperturbed Hamiltonian with $\phi^x=0$ are used to define the states of the computational basis:
\bea
\ket{\!\up~\!\!\!(s)} &\equiv &\frac{1}{\sqrt{2}}\left( \ket{1(s)} + \ket{0(s)}  \right){|_{\phi^x=0}} \nonumber \\
\ket{\!\down~\!\!\!(s)} &\equiv &\frac{1}{\sqrt{2}}\left( \ket{1(s)} - \ket{0(s)}  \right){|_{\phi^x=0}}\,.
\label{eq:comp_bas}
\eea
These symmetric and antisymmetric combinations have opposite and well-defined circulating persistent current along the whole anneal, with the persistent current operator defined as $I_\text{p} = E_\text{L} \phi$. This justifies the use of $\ket{\!\up~\!\!\!(s)}$ and $\ket{\!\down~\!\!\!(s)}$ as states of the computational basis. 

\begin{figure*}[t]
\subfigure[]{\includegraphics[width=1\columnwidth]{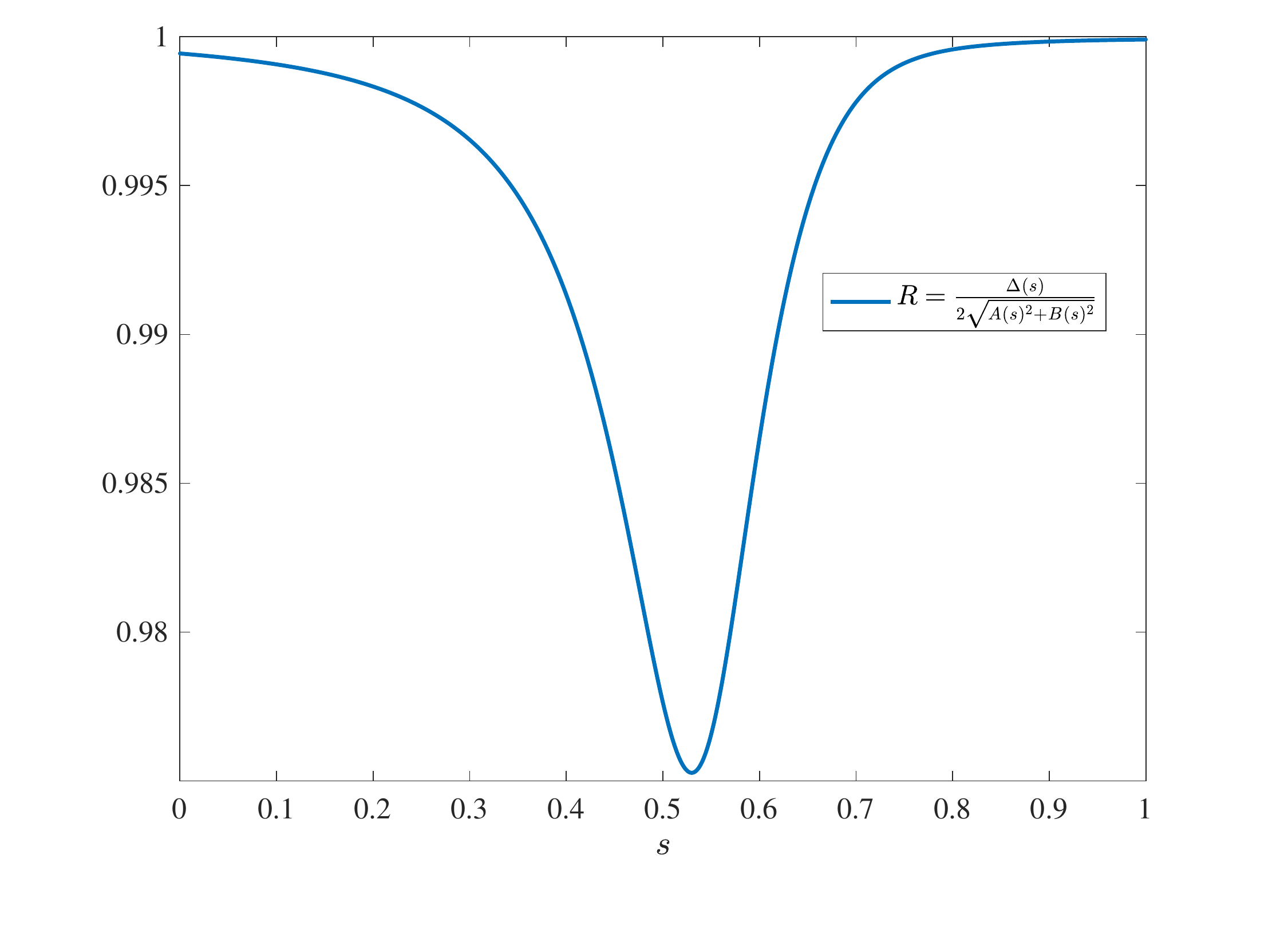} \label{fig:6a}}
\subfigure[] {\includegraphics[width=1\columnwidth]{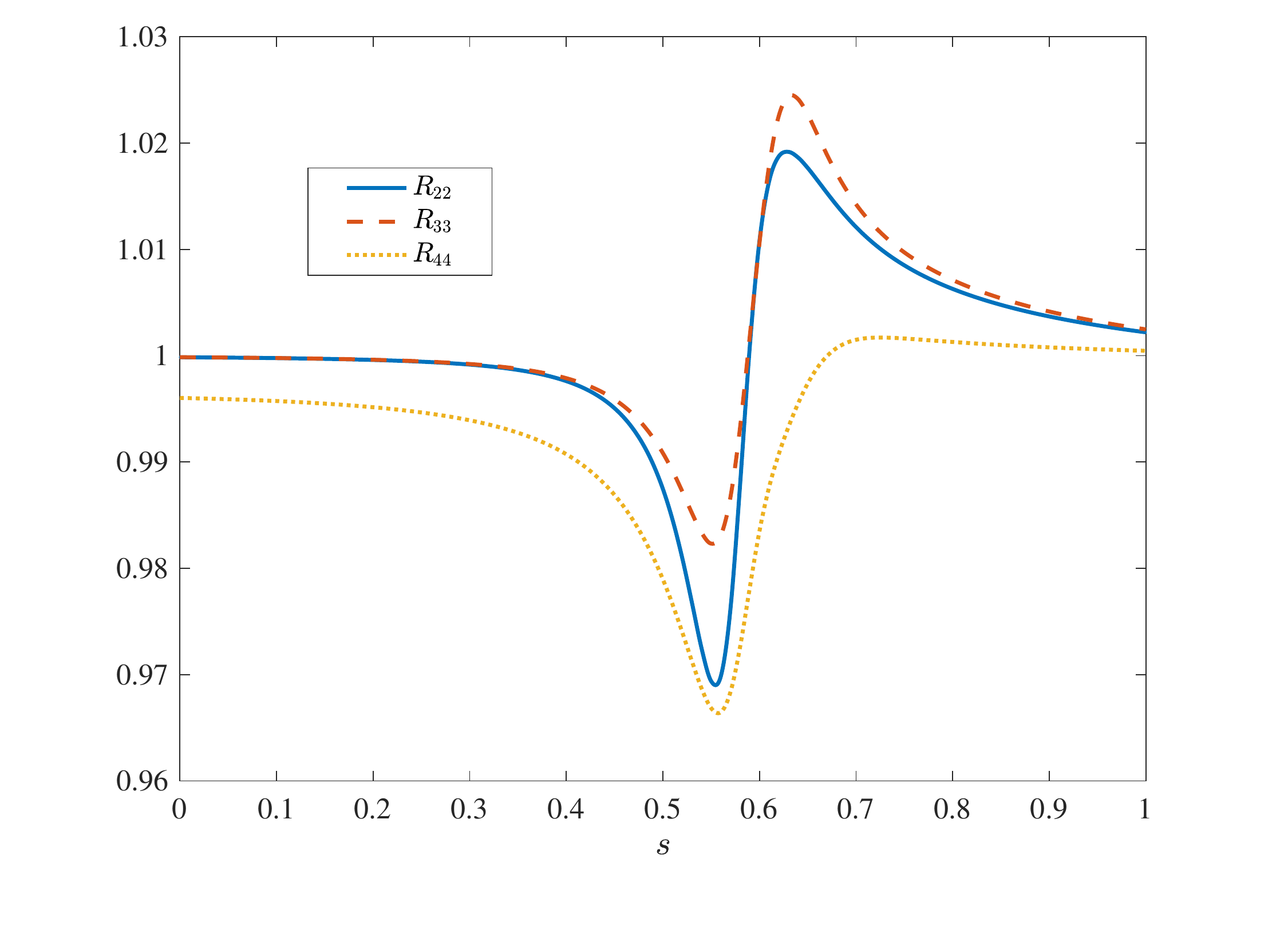} \label{fig:6b}}
\caption{(a) One C-shunt flux-qubit with: ratio between the exact gap $\Delta(s)$ and the gap estimated via the perturbative approach described in section~\ref{app:E} used to define the computational basis along the anneal and compute the profile functions $A(s)$ and $B(s)$. (b) Two coupled CJJ flux-qubits with $h_1=1,h_2=0.4,J_{12}=-0.7$: similar ratios between the exact  gaps and the gaps computed with the Hamiltonian written in the computational basis. Note that the exact gaps are well approximated by the mapping to the computational basis described in section~\ref{app:E}. The difference is largest towards the middle of the anneal, where the first order approximation in the control flux $\phi^x$ is is less accurate.}
\label{fig:6} 
\end{figure*}

We can now expand the flux qubit Hamiltonian $H_1$ in Eq.~\eqref{eq:H_one_qubit_DW} to first order in  $\phi^x$ to get:
\beq
H_1(s,\phi^x) = H_1(s,\phi^x=0) +  \phi^x(s) I_\text{p} + \mathcal O[ (\phi^x)^2]\,.
\eeq
Evaluating the Hamiltonian above in the basis $(\ket{\!\up~\!\!\!(s)},\ket{\!\down~\!\!\!(s)})$ then gives (up to a term proportional to identity) the Hamiltonian:
\beq
H_1 = A(s) \sigma^x +B(s)\sigma^z\,,
\eeq
where
\begin{align}
A(s) &\equiv   \bra{\up~\!\!\!(s)}  H_1(s,\phi^x=0) \ket{\!\down~\!\!\!(s)}   \nonumber \\
B(s) &\equiv    \bra{\up~\!\!\!(s)}   \phi^x(s)  I_\text{p} \ket{\!\up~\!\!\!(s)} \equiv \phi^x(s)  I_\text{p}(s)  \,.
\end{align}
The profile functions above are completely controlled via the external fluxes $ \phi_{\text{CJJ}}(s)$ and $\phi^x(s)$. 

The schedule of the flux $\phi^x(s)$ and thus of the profile function $B(s)$ is further constrained in the case of multi-qubit interactions. The interaction potential is given by Eq.~\eqref{eq:H_int}, whose expectation value in the computational basis is given by:
\beq
  \bra{\up_1~\!\!\!(s)~\!\!\!\up_2~\!\!\!(s)}   P_{\text{int}}  \ket{~\!\!\!\up_1~\!\!\!(s)~\!\!\!\up_2~\!\!\!(s)} \simeq -J_{12}  E_\text{M} E_\text{L}^{-2} I_\text{p}(s)^2\,,
\eeq
from which it follows that the interaction term in the effective Hamiltonian is given by $H_{\text{int}} =-J_{12}  E_\text{M} E_\text{L}^{-2} I_\text{p}(s)^2 \sigma_1^z\sigma_2^z$. To ensure the same annealing schedule for the local and interaction terms, the control field $\phi^x$ is then chosen to be proportional to the persistent current $\phi^x(s)=  E_\text{M} E_\text{L}^{-2} I_\text{p}(s)$. This implies that $B(s) =   E_\text{M} E_\text{L}^{-2} I_\text{p}(s)^2$. 

We have considered an annealing schedule linear in the control field  $ \phi_{\text{CJJ}}(s)$ [see Fig.~\ref{fig:4a}]. The corresponding values for the control flux $\phi^x(s)$ and profile functions $A(s)$ and $B(s)$ are shown in Fig.~\ref{fig:4b} and are computed using the prescription described in this section, using the numerical methods of the next section. Figure~\ref{fig:6} shows the ratios between the exact gaps computed by numerical diagonalization of Eqs.~\eqref{eq:H_one_qubit} and \eqref{eq:H_two_qubit}, and the gaps computed by diagonalizing the Hamiltonians Eqs.~\eqref{eq:H_1_rot} and \eqref{eq:H_2_rot} (without geometric terms), when the profile functions $A(s)$ and $B(s)$ are computed as described in this section. Due to the perturbative expansion, the method described in this section only approximately recovers the exact gaps.  The relative  error is largest (up to 2\%) in the middle of the anneal, when the gaps close.

\subsubsection{C-shunt Flux-Qubit}

As in the CJJ case, we will treat $\phi^x$ as a small perturbation. Expanding the Hamiltonian \eqref{eq:H_one_qubit} to first order in $\phi^x$ gives: 
\beq
H_1(\phi^x) = H_1(\phi^x = 0)  +  2 E_\text{J}\phi^x \cos(\phi^x_{\text{CJJ}}/2)\sin(2 \phi)\,.
\eeq
The Hamiltonian above has the following representation:
\beq
H_1 = A(s) \sigma^x +B(s)\sigma^z\,,
\eeq
with 
\begin{align}
A(s) &=   \left(\bra{\up~\!\!\!(s)}  H(s,\phi^x=0) \ket{\!\down~\!\!\!(s)} \right.  \nonumber \\
B(s) &=  2 E_\text{J} \phi^x \cos[\phi^x_{\text{CJJ}}(s)/2]   \bra{\up~\!\!\!(s)}   \sin(2 \phi_m) \ket{\!\up~\!\!\!(s)} \equiv \nonumber \\
 & \equiv  \phi^x \mathcal I_\text{p}(s)\,.
\end{align}
For consistency with the CJJ example, we have taken $\phi^x(s) \equiv  E_{\text{M}}E_{\text{L}}^{-2} I_\text{p}(s)^2/\mathcal I_\text{p}(s)\ $ with $ E_{\text{M}}^{-1}E_{\text{L}}^2 = 10^{4}$GHz, such that $B(s)$ is proportional to the square of the persistent current.

\section{Proof that $G$ transforms as a geometric connection}
\label{app:B}

\begin{figure*}[t]
\subfigure[\, ]{\includegraphics[width=1\columnwidth]{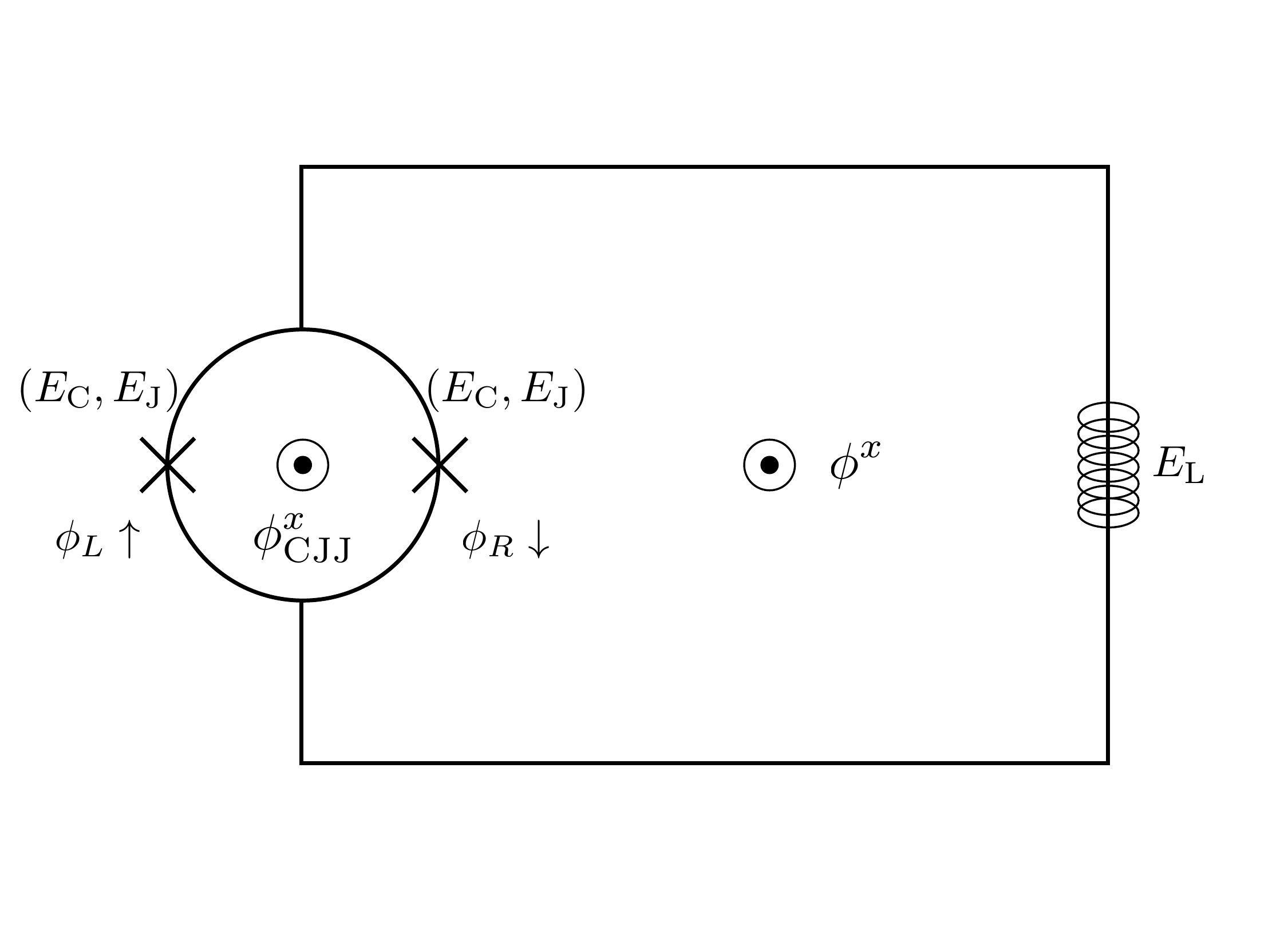} \label{fig:7a}}
\subfigure[\, ] {\includegraphics[width=1\columnwidth]{fig7a} \label{fig:7b}}
\caption{(a) Compound Josephson Junction flux-qubit. (b) Capacitively shunted flux-qubit.}
\label{fig:7} 
\end{figure*}

Let us show that $G$ transforms as in Eq.~\eqref{eq:tildeG'}, i.e., that $ G^{\text{C}}(s) = V(s) G(s) V^\dagger(s) + i V(s) \dot{V}^\dagger(s)$.

Recall that we transform from the basis $\ket{a(s)}$ (e.g., the energy eigenbasis) to the new basis $\ket{\psi^{\text{C}}_a(s)}$ (e.g., the computational basis) using the unitary $V(s)$. Thus, $\ket{a(s)} =  \sum_b V_{ba}(s)\ket{\psi^{\text{C}}_b(s)}  $. From now on we drop the explicit $s$-dependence for simplicity.
By definition, $
G^{\text{C}}_{ab} = \bra{\psi^{\text{C}}_a}G^{\text{C}}\ket{\psi^{\text{C}}_b} = i\braket{\psi^{\text{C}}_a}{\dot{\psi}^{\text{C}}_b}$. Using these two expressions, we have:
\bes
\begin{align}
G_{ab} & = \bra{a}G\ket{b} = i  \braket{a}{\dot{b}} =  \\
&= i\sum_{cd}\bra{\psi^{\text{C}}_c(s)}V^\dagger_{ac}\left[ \dot{V}_{bd}\ket{\psi^{\text{C}}_b(s)} + {V}_{bd}\ket{\dot{\psi}^{\text{C}}_b(s)}\right]\\
&= i\sum_{d}V^\dagger_{ab} \dot{V}_{bd} + \sum_{cd} V^\dagger_{ac}G^{\text{C}}_{cb}{V}_{bd}\\
&=i(V^\dagger\dot{V})_{ab} + (V^\dagger G^{\text{C}}V)_{ab}\ ,
\end{align}
\ees
i.e., $G^{\text{C}} = VGV^\dagger - i\dot{V}V^\dagger$, 
which gives the desired result after using unitarity to write $\dot{V}{V}^\dagger = -V \dot{V}^\dagger$.

\section{Flux-Qubit Hamiltonians}
\label{app:D}

The Langrangian of a superconducting circuit with Josephson junctions is generically written as follows \cite{RevModPhys.73.357}:
\beq
\mathcal L = \frac12  \sum_c  E_{\text{C},c}^{-1} \dot \phi^2_c  + \sum_j E_{\text{J},j} \cos \phi_j - \sum_l \frac12 E_{\text{L},l}{(\phi_l-\phi_l^x)^2}\,,
\eeq
where the first term is the kinetic term, the second is the junction energy, and the third is the induction energy. The $\phi_c$ are the phase differences at each capacitance. The charging energy is $E_{\text{C},c}  \equiv {(2 e)^2}/{ C_c}$, where $C_c$ is the $c$-th capacitance and $e$ the electron charge. The junction energies  are given by $E_{\text{J},j}  \equiv I_{c,j} (\Phi_0/{ 2 \pi})$, where $I_{c,j}$ is the critical current of the $j$-th junction. The induction energies are given by  $E_{\text{L},l} \equiv (\Phi_0/{ 2 \pi})^2/L_l$, where $L_l$ is the induction of the $l$-th loop. The conjugate momenta are $p_c = \partial \mathcal L/ \partial {\dot{\phi}_c} =  E_{C,c}^{-1}\dot \phi_c$. The circuit is quantized by promoting the momenta to operators: $p_c \mapsto - i \partial_{\phi_c} $. The corresponding Hamiltonian is:
\beq
H = - \sum_c  \frac{E_{\text{C},c}}{2} \partial_{\phi_c}^2  - \sum_j E_{\text{J},j} \cos \phi_j + \sum_l \frac{E_{\text{L},l}}{2}{(\phi_l-\phi_l^x)^2}\, .\nonumber
\eeq

\subsubsection{Compound Josephson Junction (CJJ) Flux-Qubit}

The basic design of a CJJ flux-qubit~\cite{yan2016flux} is shown in Fig.~\ref{fig:7a}. We assume the same charging and junction energies  for the two Josephson junctions and 
a negligible  inductance of the small loop. The Hamiltonian for this device can then be written  as:

\bea
H_{\text{CJJ}} &=& - \frac12 E_\text{C} \left(  \partial^2_{ \phi_L} +  \partial^2_{ \phi_R} \right) - E_\text{J}\left(\cos\phi_L+\cos\phi_R \right) +\nonumber \\ 
 & + & \frac{1}{2}E_\text{L}[(\phi_L-\phi_R)/2-\phi^x]^2\,,  
 \label{eq:D3}
 \eea
i.e.,  the sum of the charging, junction and induction energies of the circuit.  By defining $\phi \equiv  (\phi_L-\phi_R)/2$ and $\phi_{\text{CJJ}}= \phi_R+\phi_L$ we have  $ \partial_{ \phi_{L,R}} =  \partial_{\phi_{\text{CJJ}}}\pm1/2 \partial_{\phi}$  , from which we obtain:
\bea
H_{\text{CJJ}} &=& -\frac12 \frac{E_\text{C}}{2} \partial^2_{ \phi}  - 2 E_\text{J}\cos\left(\frac{\phi^x_\text{CJJ}}{2}\right)\cos\phi  + E_\text{L}\frac{(\phi-\phi^x)^2}{2}\, . \nonumber \\
 \eea
where we have neglected the term $ -\frac{1}{2} (2E_\text{C}) \partial^2_{ \phi_\text{CJJ}}$ since the small loop inductance gives $\phi^x_{\text{CJJ}}= \phi_{\text{CJJ}}$, i.e., the flux $\phi_{\text{CJJ}}$ is locked to the external flux $\phi^x_{\text{CJJ}}$. The equation above reduces to Eq.~\eqref{eq:H_one_qubit_DW} with the redefinition $\phi^x_\text{CJJ} \mapsto 2\pi -\phi^x_\text{CJJ}$.

\subsubsection{Capacitively Shunted (C-shunt) Flux-Qubit}

The basic design of a C-shunt flux-qubit~\cite{orlando1999superconducting,yan2016flux} involves four Josephson junctions  and a large shunting capacitance and is shown in Fig.~\ref{fig:7b}. We start by writing the kinetic term:
\beq
K = \frac12  E_{\text{C}}^{-1}  \left( \dot \phi^2_1 + \dot \phi^2_2 + \dot \phi_{L}^2 + \dot \phi_R^2     \right) + \frac12 E^{-1}_{\text{S}} \left(   \dot \phi_1 - \dot \phi_2 \right)^2\, , \\
\label{eq:K}
 \eeq
where the first term comes from the junctions while the last term is the shunting capacitor energy with $E_{\text{S}} = {(2e)^2}/{C_\text{S}}$. 
By defining  $\phi_{\text{CJJ}}= \phi_R+\phi_L$,  $\phi= (\phi_L-\phi_R)/2+\phi_2-\phi_1$ and  $\phi_{\pm} = (\phi_1\pm \phi_2)/2$ we get $\dot \phi_{1,2} = \dot \phi_{+}\pm  \dot \phi_{-}$ and $\dot \phi_R = -\dot \phi_L = -2 \dot \phi_{-}$, where we have set $\dot \phi_{\text{CJJ}} =  \dot \phi = 0$, i.e., the fluxes  $ \phi_{\text{CJJ}}$  and  $\phi$ are constant and locked to the external fluxes ($\phi_{\text{CJJ}}=\phi^x_{\text{CJJ}}$,  $\phi = \phi^x$) due to the small inductances. We can then rewrite $K$ in Eq.~\eqref{eq:K} as 
\bea
K &=&  \frac12  E_{\text{C}}^{-1} \left( 2 \dot \phi^2_+ +  10 \dot \phi^2_-   \right)+ \frac12 \left( \frac{E_\text{S}}{4} \right)^{-1} \dot \phi^2_-  \simeq  \nonumber \\
& \simeq &  \frac12 \left( \frac{E_\text{S}}{4} \right)^{-1} \dot \phi^2_-\, ,
\eea
where in the last step we neglected the light (high-frequency) mode $\phi_+$  ($E_{\text{S}} \ll E_{\text{C}}$ due to the large shunting capacitance). The potential term due to the Josephson junctions is just the sum of the four junction energies:
\bea
P &=&  - E_\text{J}\left(\cos\phi_1+\cos\phi_2 + \cos\phi_L+\cos\phi_R \right)   \\
&=&- 2 E_\text{J}\left[  \cos\phi_-\cos \phi_+ + \cos(\phi^x_{\text{CJJ}}/2)\cos(\phi^x+2 \phi_-)  \right]\,. \nonumber
\eea
We then get the final Hamiltonian for the C-shunt qubit:
\bea
\label{eq:c-shunt}
H &=& -   \frac12 \left(\frac{E_\text{S}}{4} \right) \partial^2_{\phi_-} +  \\
&-& 2 E_\text{J}\left[  \cos\phi_-\cos \phi_+ + \cos(\phi^x_{\text{CJJ}}/2)\cos(\phi^x+2 \phi_-)  \right]\,, \nonumber
\eea
which reduces to Eq.~\eqref{eq:H_one_qubit} after we set $\phi_+ = 0$ (i.e. the value that minimize the potential), and rename $\phi_- \mapsto \phi$.

\section{Derivation of Eq.~\eqref{eq:18}}
\label{app:C}

The effective Hamiltonian Eq.~\eqref{eq:H_1}  $H_1^{\text{eff}}(s) =  t_f\frac{\Delta(s)}{2} \sigma^z -g(s)  \sigma^y$
is written in the basis defined by the instantaneous energy eigenbasis. A more conventional choice is to rewrite the Hamiltonian above in the computational basis, as in Eq.~\eqref{eq:H_1_rot}. This can be done via a unitary transformation of the form
\beq
V(s) = \exp \left[ i \theta(s) \,\sigma^y \right]\,,
\eeq
with $\theta(s)$ to be determined. Now,
\bea
V(s)  \sigma^z  V^\dagger(s) = -\sin [2 \theta(s)]  \sigma^x   +\cos [2 \theta(s)] \sigma^z  \,,
\eea
from which we find, using Eq.~\eqref{eq:comp_basis}:
\beq
A(s) = \frac{\Delta(s)}{2}  \sin [2 \theta(s)]\,, \quad B(s) = \frac{\Delta(s)}{2}  \cos [2 \theta(s)]\ ,
\eeq
from which Eq.~\eqref{eq:Delta-A-B} follows.  For the geometric term we use Eq.~\eqref{eq:tildeG'} to get
\begin{align}
g^y(s) \sigma^y &\equiv V(s) g(s)\sigma^y  V^\dagger(s)  + i V(s) \dot{V}^\dagger(s) \nonumber \\
&= \left[g(s) + \dot{\theta}(s)  \right] \sigma^y \,. 
\end{align}
Since $\theta(s) = \arctan[A(s)/B(s)]/2$, we have:
\beq
\dot{\theta}(s) = \frac{2}{\Delta^2(s)}\left[\dot{A}(s)B(s)  -  A(s) \dot{B}(s) \right]\,. 
\eeq
Combining the last two equations yields Eq.~\eqref{eq:g^y} reported in the main text.

\section{Numerical Methodology}
\label{app:F}

All the ``static" quantities, e.g., the quantities shown in Figs.~\ref{fig:1a}, \ref{fig:2a}, \ref{fig:4}, \ref{fig:5} and \ref{fig:6}, are determined at a given value of the schedule parameter $s$ by first numerically computing the wave functions $\ket{0(s)}$ and $\ket{1(s)}$ [see, e.g., Fig.~\ref{fig:4} and Fig.~\ref{fig:5}]. We first discretized the flux-qubit Hamiltonian Eq.~\eqref{eq:H_ann}. For example, the one-qubit Hamiltonian Eq.~\eqref{eq:H_1_LL} is reduced to the following $L$-dimensional system:
\bea
 &H_1^{\text{discr}}({\phi},s,h) \approx - \frac{E_\text{C}}{8} \frac{(L-1)^2}{(\phi_L-\phi_{-L})^2}\left(
\begin{array}{cccc}
  -2 &   1 &    &\\
  1& \ddots   & \ddots  &  \\
  &   \ddots& \ddots   & 1 \\
  &    & 1   & -2
\end{array}
\right)+& \nonumber \\
&+ \left(
\begin{array}{cccc}
  P(\phi_{0},s,h) &    &    &\\
  & P(\phi_{1},s,h)   &   &  \\
  &   &  \ddots  &  \\
  &    &    & P(\phi_{L-1},s,h)
\end{array}
\right)\,,& \nonumber
\eea
where the first term is the discretized Laplacian and the continuous flux $\phi$ is discretized as $\phi_i = \phi_{-L}+ i (\phi_L-\phi_{-L})/(L-1)$, $i = 0,\dots,L-1$, with $L$ being the size of the mesh. Similarly we can discretize the two-qubit Hamiltonian  Eq.~\eqref{eq:H_two_qubit} to obtain an  $L^2$-dimensional system. We used $L = 600$, which was sufficient for  numerical convergence. We numerically computed all the static functions on a mesh of $100$ points for the annealing parameter $s$. The value of all functions at all other intermediate points, when required, where computed via a cubic interpolation.

Once all the static quantities where computed, the ``dynamic" quantities of Figs.~\ref{fig:1b}, \ref{fig:2b} and \ref{fig:3} were computed by solving the Schr{\"o}dinger equations resulting from the effective one- and two-qubit Hamiltonians Eqs.~\eqref{eq:H_1}, \eqref{eq:H_1_rot} and \eqref{eq:H_2_rot}. We used a standard \texttt{ode45} solver provided with Matlab.



\bibliography{refs}

\end{document}